\documentclass[acmsmall,screen]{acmart}

\newif\ifExtended
\Extendedtrue

\AtBeginDocument{%
	\providecommand\BibTeX{{%
			\normalfont B\kern-0.5em{\scshape i\kern-0.25em b}\kern-0.8em\TeX}}}

\setcopyright{rightsretained}
\acmJournal{PACMPL}
\acmYear{2021}
\acmVolume{5}
\acmNumber{OOPSLA}
\acmArticle{99}
\acmMonth{10}
\acmPrice{}
\acmDOI{10.1145/3485476}
\copyrightyear{2021}
\acmSubmissionID{oopsla21main-p14-p}

\usepackage{booktabs}   
\usepackage{subcaption} 

\usepackage[capitalize]{cleveref}

\usepackage{amsmath,mathtools}
\usepackage{proof}
\usepackage{mathpartir}
\usepackage{xspace}
\usepackage{paracol}
\usepackage{lipsum}
\usepackage{xcolor}
\usepackage{blindtext}
\usepackage{color}
\usepackage[toc,page]{appendix}

\usepackage{setspace}
\usepackage{cuted}

\graphicspath{{./figs/}}

\newcommand{\FIXME}[1]{\colorbox{yellow}{\textbf{FIXME:}} \textbf{#1}}

\newcommand{\pengbo}[1]{\relax} 

\newcommand{\Triple}[6]{\left[#1\right],\ \{#2\}\ #3,\ #4\ \{#5\},\ \left[#6\right]}
\newcommand{\TripleAH}[5]{\left[#1\right],\ \{#2\}\ #3\ \{#4\},\ \left[#5\right]}
\newcommand{\TripleNA}[4]{\{#1\}\ #2,\ #3\ \{#4\}}
\newcommand{\TripleA}[3]{\{#1\}\ #2\ \{#3\}}
\newcommand{\nat}{\mathbb{N}}
\newcommand{\PQ}{\textcolor{blue}}
\newcommand{\rPQ}{\textcolor{red}}
\newcommand{\Lock}[2]{\mathit{Lock}(#1,#2)}

\newcommand{\na}{\textbf{na}}
\newcommand{\rel}{\textbf{rel}}
\newcommand{\acq}{\textbf{acq}}
\newcommand{\rlx}{\textbf{rlx}}
\newcommand{\scx}{\textbf{sc}}
\newcommand{\relacq}{\textbf{rel\_acq}}
\newcommand{\low}{\textbf{low}}
\newcommand{\high}{\textbf{high}}

\newcommand{\A}{\mathcal{A}}

\newcommand{\sem}[1]{\llbracket #1 \rrbracket}
\newcommand{\hoare}[3]{\{ #1 \}~#2~\{ #3 \}}

\newcommand{\True}{\texttt{true}}
\newcommand{\False}{\texttt{false}}
\newcommand{\emp}{\mathsf{emp}}
\newcommand{\sep}{\mathrel{\star}}

\newcommand{\cmd}[1]{\texttt{#1}}

\newcommand{\IF}{\cmd{if}}
\newcommand{\THEN}{\cmd{then}}
\newcommand{\ELSE}{\cmd{else}}

\newcommand{\Let}[3]{\mathbf{let}\ #1 = #2\ \mathbf{in}\ #3}
\newcommand{\If}[3]{\mathbf{if}\ #1\ \mathbf{then}\ #2\ \mathbf{else}\ #3}
\newcommand{\Repeat}[1]{\mathbf{repeat}\ #1\ \mathbf{end}}
\newcommand{\Parallel}[2]{#1 \| #2}
\newcommand{\Alloc}[1]{\mathbf{alloc}_{#1}()}
\newcommand{\Load}[2]{[#1]_{#2}}
\newcommand{\Store}[3]{[#1]_{#2} := #3}
\newcommand{\CAS}[5]{\mathbf{CAS}_{#1,#2}(#3,#4,#5)}

\newcommand{\Alab}[2]{\mathrm{A}_{#1}(#2)}
\newcommand{\skiplab}{\mathrm{skip}}
\newcommand{\Wlab}[3]{\mathrm{W}_{#1}(#2,#3)}
\newcommand{\Rlab}[3]{\mathrm{R}_{#1}(#2,#3)}
\newcommand{\RMWlab}[4]{\mathrm{RMW}_{#1}(#2,#3,#4)}

\newcommand{\union}{\cup}
\newcommand{\inter}{\cap}

\newcommand{\Val}{\mathsf{Val}}
\newcommand{\Var}{\mathsf{Var}}
\newcommand{\Loc}{\mathsf{Loc}}
\newcommand{\Exp}{\mathsf{Exp}}
\newcommand{\AExp}{\mathsf{AExp}}
\newcommand{\Act}{\mathsf{Act}}
\newcommand{\AName}{\mathsf{AName}}
\newcommand{\sbo}{\mathit{sb}}
\newcommand{\rf}{\mathit{rf}}
\newcommand{\mo}{\mathit{mo}}
\newcommand{\hb}{\mathit{hb}}
\newcommand{\sw}{\mathit{sw}}
\newcommand{\lab}{\mathit{lab}}
\newcommand{\sco}{\mathit{sc}} 
\newcommand{\res}{\mathit{res}}
\newcommand{\lst}{\mathit{lst}}
\newcommand{\fst}{\mathit{fst}}

\newcommand{\rexec}[7]{\langle #1, #2, #3, #4, #5, #6, #7 \rangle}
\newcommand{\exec}[6]{\langle #1, #2, #3, #4, #5, #6 \rangle}
\newcommand{\Consistent}[1]{\mathsf{Consistent}(#1)}

\newcommand{\Afalse}{\mathsf{false}}
\newcommand{\Aemp}{\mathsf{emp}}
\newcommand{\Aimplies}[2]{#1 \Rightarrow #2}
\newcommand{\Aforall}[3]{\forall (#1,#2), #3}
\newcommand{\Apt}[4]{(#1,#2) \mapsto (#3,#4)}
\newcommand{\Aptu}[2]{\mathsf{Uninit}(#1,#2)}
\newcommand{\Astar}[2]{#1 \star #2}
\newcommand{\Ainit}[1]{\mathsf{Init}(#1)}
\newcommand{\Alow}[2]{\mathsf{Low}(#1,#2)}
\newcommand{\Arel}[2]{\mathsf{Rel}(#1,#2)}
\newcommand{\Aacq}[2]{\mathsf{Acq}(#1,#2)}
\newcommand{\Armw}[2]{\mathsf{RMWAcq}(#1,#2)}
\newcommand{\Aexists}[3]{\exists (#1,#2), #3}

\newcommand{\Assn}{\mathsf{Assn}}
\newcommand{\Atrue}{\mathsf{true}}
\newcommand{\Precise}[1]{\mathsf{precise}(#1)}

\newcommand{\undefine}{\mathsf{undef}}
\newcommand{\uninit}{\mathbb{U}}
\newcommand{\Ffalse}{\mathsf{False}}
\newcommand{\Femp}{\mathsf{Emp}}
\newcommand{\NA}[1]{\mathsf{NA}\left[ #1 \right]}
\newcommand{\mapsNA}[2]{#1 \mapsto \NA{#2}}
\newcommand{\hA}[4]{\mathsf{Atom}\left[ #1, #2, #3, #4 \right]}
\newcommand{\mapsA}[5]{#1 \mapsto \hA{#2}{#3}{#4}{#5}}
\newcommand{\hmap}{\mathit{hmap}}
\newcommand{\HS}{\mathit{H}}
\newcommand{\V}{\mathit{V}}
\newcommand{\Valid}[6]{\mathsf{Valid}(#1,#2,#3,#4,#5,#6)}
\newcommand{\Heap}{\mathsf{Heap}_{\mathsf{spec}}}
\newcommand{\isAcq}[1]{\mathsf{isAcq}(#1)}
\newcommand{\isRel}[1]{\mathsf{isRel}(#1)}
\newcommand{\isNA}[1]{\mathsf{isNA}(#1)}

\newcommand{\metaITE}[3]{\mathit{if}\ #1\ \mathit{then}\ #2\ \mathit{else}\ #3} 
\newcommand{\setof}[1]{[\![#1]\!]}

\newcommand{\CCof}[1]{\mathit{CC}\setof{#1}}
\newcommand{\SBout}[2]{\mathsf{SBout}_{#1}(#2)}
\newcommand{\SBin}[2]{\mathsf{SBin}_{#1}(#2)}
\newcommand{\SWout}[2]{\mathsf{SWout}_{#1}(#2)}
\newcommand{\SWin}[2]{\mathsf{SWin}_{#1}(#2)}
\newcommand{\Pre}[2]{\mathsf{Pre}_{#1}(#2)}
\newcommand{\functo}[2]{#1\rightarrow #2}
\newcommand{\Pcons}[6]{\mathsf{PartialConsistent}(#1, #2, #3, #4, #5, #6)}
\newcommand{\Resp}[2]{\mathsf{Resp}_{#1}(#2)}
\newcommand{\simsafe}[6]{\mathsf{safe}^{#1}_\chi (\res,#2,#3,#4,#5,#6,\A_{prg},\A_{ctx},\fst,\lst,E',Q,HQ)}

\newcommand{\CLow}[1]{\mathsf{ClassLow}(#1)}
\newcommand{\CHigh}[1]{\mathsf{ClassHigh}(#1)}
\newcommand{\Mod}[2]{(#1\ \mathbf{mod}\ #2)}
\newcommand{\Sender}[1]{\mathsf{Sender}(#1)}
\newcommand{\Recver}[1]{\mathsf{Recver}(#1)}
\newcommand{\SenderC}{\mathsf{Sender}}
\newcommand{\RecverC}{\mathsf{Recver}}
\newcommand{\preC}[2]{\left[#1\right],\ \{#2\}}
\newcommand{\postC}[2]{\{#2\},\ \left[#1\right]}
\newcommand{\postSC}[2]{\{#2\}}
\newcommand{\LockC}[1]{\mathsf{LockMSM}(#1)}

\newcommand{\Prog}{\mathit{Prog}}
\newcommand{\MP}{\mathit{MP}} 

\newcommand{\SecCSL}{\textsc{SecCSL}\xspace}

\DeclareRobustCommand{\rchi}{{\mathpalette\irchi\relax}}
\newcommand{\irchi}[2]{\raisebox{\depth}{$#1\chi$}} 
\newcommand{\defeq}{\stackrel{\text{def}}{=}}

\newenvironment{DIFnomarkup}{}{}

\begin{document}

\title[SecRSL: Security Separation Logic for C11 Release-Acquire Concurrency]{SecRSL: Security Separation Logic for C11 Release-Acquire Concurrency \ifExtended (Extended version with technical appendices)\fi}         


\author{Pengbo Yan}
\orcid{0000-0003-0396-8343}                                    
\affiliation{
  \institution{University of Melbourne}            
  \country{Australia}                    
}
\email{pengbo.yan@unimelb.edu.au}          

\author{Toby Murray}
\orcid{0000-0002-8271-0289} 
\affiliation{
  \institution{University of Melbourne}           
  \country{Australia}                   
}
\email{toby.murray@unimelb.edu.au}         

\begin{abstract}
  We present Security Relaxed Separation Logic (SecRSL), a
  separation logic for proving information-flow security of C11 programs
  in the Release-Acquire fragment with relaxed accesses. SecRSL is the
  first security logic that (1)~supports weak-memory reasoning about
  programs in a high-level language; (2)~inherits separation logic's
  virtues of compositional, local reasoning about (3)~expressive security
  policies like value-dependent classification.

  SecRSL is also, to our knowledge, the first security logic developed
  over an axiomatic memory model. Thus we also present 
  the first definitions of information-flow security for an
  axiomatic weak memory model, against which we prove SecRSL sound.
  SecRSL ensures that programs satisfy a constant-time security guarantee,
  while being free of undefined behaviour.

  We apply SecRSL to implement and verify the
  functional correctness and constant-time
  security of a range of concurrency primitives, including
  a spinlock module, a mixed-sensitivity mutex, and
  multiple synchronous channel implementations. Empirical performance
  evaluations of the latter demonstrate SecRSL's power to
  support the development of secure and performant concurrent C programs.
\end{abstract}



\maketitle

\section{Introduction}

Logics for proving that concurrent programs do not leak sensitive information
have received much recent
study~\cite{Murray_SPR_16,Murray_SE_18,Karbyshev+:POST18,Ernst_Murray_19,schoepe2020veronica,frumin2019compositional}. A common thread of much recent work has been adapting
ideas from concurrent separation logic~\cite{OHearn_04}
to reason about secure information flow. Indeed, recent security logics
purposefully closely resemble traditional concurrent separation logics~\cite{Ernst_Murray_19,frumin2019compositional}. Besides providing a familiar, compositional
and elegant setting in which
to carry out security proofs, such logics have also proved amenable to
automated verification via symbolic execution~\cite{Ernst_Murray_19}.

Security separation logics have so far been confined to reasoning over
sequentially-consistent memory models. There exist a handful of
information-flow logics and type
systems for weak memory consistency models~\cite{vaughan2012secure,mantel2014noninterference,smith2019value}. However they lack the
local reasoning abilities, and consequent scalability, afforded by
separation logics. They also lack separation logic's support for
reasoning about
resource ownership transfer with invariants, which is crucial for expressive
reasoning. Additionally, these logics have targeted machine-level
weak memory models, and so are not readily applicable for reasoning about
programs in higher level languages. 

%
%
%
%
%

In some sense, these limitations are not surprising. Logics for
information-flow security are almost exclusively proved sound against
\emph{operational} semantic models. Yet formal weak memory models are often
specified \emph{axiomatically}, in which program behaviours are represented
as mathematical objects 
constrained by the axioms of the weak memory model.

To our knowledge there do not even exist prior definitions of
information-flow security for such models, without which one cannot even
state the soundness theorems for any proposed logic. Even if one had
such properties, another necessary
but lacking ingredient for proving the soundness of such logics is suitable
compositional, inductive properties that state the semantic meaning of
the logic's judgements.  The shape of such properties is well understood
for traditional (non-security) separation logics~\cite{Vafeiadis_Narayan_13,FSL,FSLplusplus}, many of which share very similar definitions. Yet little is
known about what they should look like for security separation logics.

In this paper, we introduce \emph{Security Relaxed Separation Logic} (SecRSL).
SecRSL is the first concurrent separation logic for
reasoning about secure information flow in a relaxed memory model. It targets
the Release-Acquire fragment of C11 with relaxed accesses (\cref{C11}),
and thus supports
reasoning about high-level programs for the first time. It also inherits
the local reasoning, expressiveness, and compositionality of
traditional concurrent separation logics.

SecRSL's design (\cref{logic})
combines ideas from two prior logics: (1)~Ernst and Murray's
Security Concurrent Separation Logic (\SecCSL)~\cite{Ernst_Murray_19}, an
information-flow security analogue of traditional concurrent separation logic
for sequential consistent concurrency; and (2)~Vafeiadis and Narayan's
Relaxed Separation Logic (RSL)~\cite{Vafeiadis_Narayan_13}. Thus SecRSL
inherits \SecCSL's support for proving expressive security policies
like those involving \emph{value-dependent classification}~\cite{Murray_SPR_16}, in which the
sensitivity of one variable can change in response to changes in the
value held by another.  It also inherits RSL's ability to reason about
ownership transfer with invariants via C11's Release-Acquire atomics.

To state the soundness of SecRSL (\cref{soundness}),
we present what is to our knowledge the
first definitions of information-flow security for an axiomatic weak memory
model (\cref{adequacy}).
We believe it should
be readily applicable to similar memory models~\cite{FSL,FSLplusplus} with
little modification, while providing guidance on how to structure such
definitions for less similar axiomatic models.
Our security definition protects against passive attackers who can
observe the contents of $\low$ (public) memory locations, as well as attackers
who can observe the program's memory access pattern (i.e.\ can see \emph{which}
locations are being accessed by the program but not the \emph{values} being
written to them). Thus SecRSL also protects against attackers who can mount
cache side-channel attacks and provides a form of
\emph{constant-time security}~\cite{barthe2019formal}. 

To prove SecRSL sound, we developed a novel, compositional definition of
relational validity (\cref{triple}) that encodes the meaning of SecRSL judgements. As with
the top-level security properties, we believe this definition is of
independent interest beyond the confines of our specific memory model,
which we inherit from RSL~\cite{Vafeiadis_Narayan_13}.

We demonstrate SecRSL's power by using it to implement and verify
the functional correctness and constant-time
security of a range of concurrency primitives (\cref{applications}).
Specifically, we demonstrate 
SecRSL's support for traditional relaxed memory reasoning
by showing how Vafeiadis and Narayan's RSL proof for a spinlock module
can be replayed in SecRSL. Thus we prove it not only functionally
correct but also to satisfy SecRSL's constant-time security guarantee.
We then extend that example to implement a \emph{mixed-sensitivity mutex}
for protecting access to data of varying sensitivity. Finally, we
implement and verify multiple implementations of a \emph{synchronous channel}
abstraction for transmitting data of varying sensitivity. We benchmark
their performance against an (unverified) sequentially-consistent
implementation. In doing so,
we find that SecRSL enables significant performance gains of up to $\sim$90\%
depending on platform. Thus demonstrating
SecRSL's power to
support the development of secure and performant concurrent C programming
abstractions.

All formal results in this paper have been mechanised in the Coq theorem
prover. The theories are available as supplementary material~\cite{SupplementaryMaterial}.

\section{Overview}\label{sec:overview}

\paragraph{C11 Release-Acquire atomics}

\begin{figure}
    \begin{center}
      $\MP_1(x) \defeq \ \ \ \ \ \ \ \ \ \ \ \ \ \ \ \ \ \ \ \ \ \ \ \ \ \ \ \ \ \ \ \ \ \ \ \ \ \ \ \ \ \ \ \ \ \ \ \ \ \ \ \ \ \ \ \ \ \ \ \ $\\
    $\Let{b}{\Alloc{\high}}{}$\\
    $\Let{a}{\Alloc{\low}}{}$\\
    $\Let{c}{\Alloc{\low}}{}$\\      
    $\Store{a}{\rlx}{0};$\\
    \qquad \begin{tabular}{l@{\quad}||@{\quad}l}
                $\Store{b}{\na}{x};$& $\Repeat{\Load{a}{\acq}}{}$\\
                $\Store{a}{\rel}{1};$& $\Let{y}{\Load{b}{\na}}{}$\\
                                     & $\Store{c}{\na}{y}$
    \end{tabular}
\end{center}
  \caption{A message-passing program\label{fig:mp}}
\end{figure}

The program~$\MP_1$
in \cref{fig:mp} is a slight adaptation of the classic \emph{message-passing}
program, and makes use of C11's Release-Acquire atomics. This program
has access to some local variable~$x$. It creates two threads and uses the 
location~$b$ to send a message containing~$x$ from the left thread to the
right one, where the location~$a$ is used to synchronise the two threads. 
The message-passing protocol initialises
$a$ to hold 0. Once~$b$ holds the message to be transferred, $a$ is updated
to hold 1 to signal to the right thread that the message (stored in~$b$)
is now ready to be read. The right thread busywaits until~$a$ is non-zero
(the \textbf{repeat} command executes its body until it returns a non-zero
value). The right thread
then loads the message from~$b$ before writing it to the location~$c$.

In the C11 Release-Acquire fragment we consider, loads and stores to memory
locations are annotated with various modes. These affect the potential
observable reorderings that are possible under the memory model.
The left thread's
store of 1 to location~$a$ carries the \emph{release} mode ($\rel$), while
the right thread's load of~$a$ (inside the busyloop) carries the
\emph{acquire} mode ($\acq$). Together these modes guarantee that if the
acquire load reads the value~$1$, then the prior store of $x$ to location~$b$
must have also completed. Hence, the subsequent load from~$b$ must read
the value~$x$ and so~$x$ will be written to location~$c$.

In SecRSL (as in RSL before it) two kinds of locations are distinguished:
\emph{atomic} locations and \emph{non-atomic} locations. Modes like
acquire and release can be used only with atomic locations; non-atomic
locations must be accessed using the non-atomic mode ($\na$). When reasoning
about the program in \cref{fig:mp} therefore, location~$a$ is treated
as an atomic
location while~$b$ and~$c$ are considered non-atomic locations.

\begin{figure}
	\includegraphics[width=80mm]{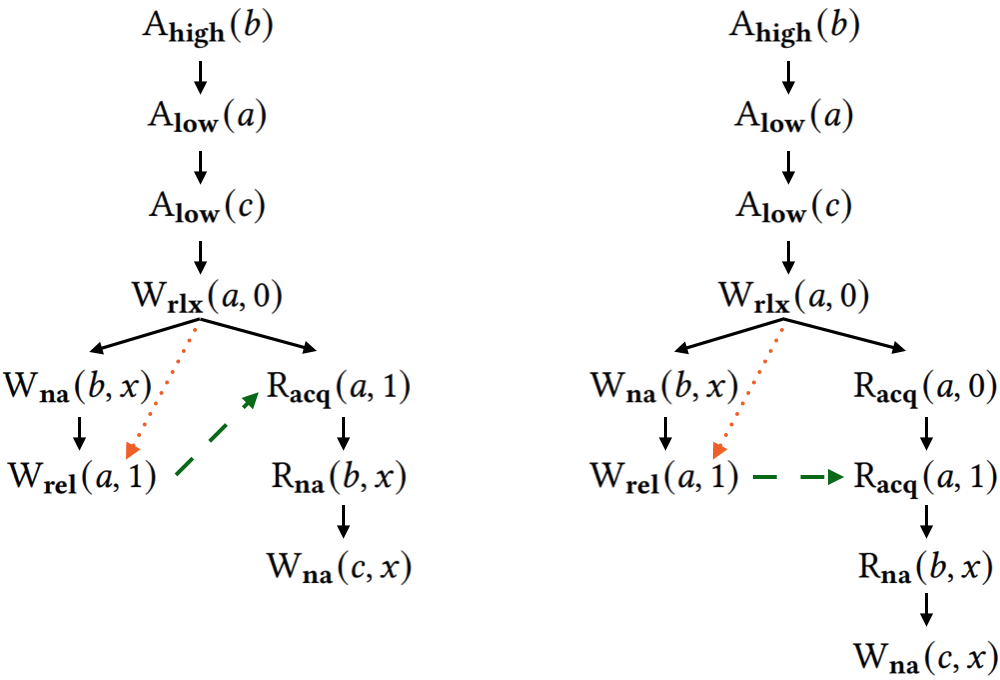}
	\caption{Two executions of the $\MP_1(x)$ program\label{fig:mpLab}. Edges depict some key relations of the memory model (\cref{sec:memory-model}):  $\sbo$ (black, solid); $\rf$ (dashed, green); $\mo$ (dotted, orange). As with much prior work, we draw the arrows for the $\rf$ relation to depict data flow: for read and write actions~$r$ and~$w$ respectively, when $\rf(r) = w$ we draw a (dashed, green) line from $w$ to $r$.}
\end{figure}

The axiomatic C11 memory model defines a program execution as a set of
actions related by several binary relations that impose ordering constraints
on those events.  Two possible executions of this program are depicted
graphically in \cref{fig:mpLab}, where the nodes depict the actions and
the edges depict various relations between them. The execution on the
left in this figure arises when the loop executes only once; the one on
the right executes the busyloop twice.

We formally define the Release-Acquire fragment of C11 and its memory
model that SecRSL
treats in \cref{C11}, which we inherit from RSL~\cite{Vafeiadis_Narayan_13}.

\paragraph{Information-flow Security}
What should information-flow security mean in this fragment of C11?
As with \SecCSL~\cite{Ernst_Murray_19}, we assume the existence
of a passive attacker who can observe certain parts of memory.
Specifically, when they are allocated
certain memory locations are deemed to be potentially attacker-observable.
Such locations might hold data that will be publicly output
or copied onto the network. We call these locations~$\low$ and
all other locations~$\high$.

Notice from the allocation statements
in \cref{fig:mp} that the atomic location~$a$ and the non-atomic
location~$c$ are both $\low$. Being conservative, atomic locations should
be considered attacker-observable, because interactions with them (e.g.~those
with~$a$ in \cref{fig:mp}) affect thread scheduling and so are
potentially observable to attackers who can measure time.

In this example, the location~$c$ is also considered attacker observable.
As is standard in information-flow security, we assume the attacker knows
the program's source code. Hence, constants that appear in the program text
are considered public.
The program only writes public constants to~$a$; however it eventually
writes~$x$ to location~$c$. Hence, the program is information-flow secure
only if~$x$ is public data.

We say that a program does not leak information (and so is information
flow secure) if all executions of that program beginning from states
that agree on the values of public (i.e.\ non-secret)
data are indistinguishable to the
attacker. Supposing the variable~$x$ holds public data initially, then
the program~$\MP_1$ is secure if for all~$x$, $x'$, if $x$ and~$x'$ are equal,
then $\MP_1(x)$ is indistinguishable from~$\MP_1(x')$.

Due to the nondeterminism inherent in axiomatic semantic models, a single
program can give rise to multiple executions: an implementation is free to
choose any one of them while remaining compliant with the semantics.
Indeed two possible behaviours of~$\MP_1(x)$ are depicted in \cref{fig:mpLab}.
This nondeterminism creates potential challenges for secure information flow.

To understand why, suppose the initial state of the
example was extended with an additional secret bit~$b$ and consider an
implementation of the weak memory semantics that chooses to produce the
left execution when $b=1$ and to produce the right one otherwise.
An attacker who can measure execution time would be able to distinguish
these two executions since one performs an extra action.

Thus, as with prior work~\cite{smith2019value},
we make the necessary assumption that the resolution
of nondeterminism in the weak memory semantics does not itself reveal
sensitive information. Since in our semantics nondeterminism arises
from the selection of memory locations returned by the \textbf{alloc}
expression, as well as from the concurrency semantics, we 
assume that neither the memory allocator nor the concurrency implementation
(the compiler, operating system scheduler, hardware, etc.) reveal secrets
to the attacker. Both assumptions are intuitively necessary in order to
proceed with source level reasoning about information-flow security.

Our threat model also includes attackers who can observe not only the
values written to $\low$ locations, but also the program's memory access
pattern: such attackers can observe which locations the program is accessing
even if they cannot see the values being written to them.

Thus as well
as guaranteeing that indistinguishable values are written to $\low$ locations,
SecRSL's top-level security property also requires that for two
executions to be indistinguishable they must access the same locations
in the same way at the same time.
Additionally its rules prevent programs from branching on secrets.
Thus it guarantees a form
\emph{constant-time security}~\cite{barthe2019formal}.

We formally define these guarantees via SecRSL's top-level security
property in \cref{adequacy}.

\paragraph{The Logic}

Recall we said the example $\MP_1$ is secure under the assumption that
its argument is public if for all arguments $x$ and~$x'$ that
are equal, $\MP_1(x)$ is indistinguishable from~$\MP_1(x')$. Thus SecRSL
is a \emph{relational} logic~\cite{Benton_04} that in general
relates the executions of two programs, and proves them indistinguishable.

SecRSL judgements have the general form\[
\Triple{HP}{P}{E}{E'}{(y,y').\ Q}{(y,y').\ HQ},
\]
where~$P$ is the precondition,
$E$ and~$E'$ are the two programs we are proving 
indistinguishable, $y$ and~$y'$ are the return values of~$E$ and~$E'$, and
$Q$ is the postcondition. $HP$ and~$HQ$ track the set of $\high$ locations
before and after the execution.

Since SecRSL is relational, its separation logic assertions relate pairs of
states~\cite{Ernst_Murray_19}. The judgement above means that if~$E$ and~$E'$
are run from starting states related by~$P$ in which all locations in~$HP$
are unobservable to the attacker, then their executions will be indistinguishable to the attacker and, moreover, $Q$ will relate the final states and
$HQ$ will track the final set of locations known to be unobservable to
the attacker.

The security of~$\MP_1$ we can thus express in SecRSL as the judgement:
$\Triple{\emptyset}{\Alow{x}{x'}}{\MP_1(x)}{\linebreak\MP_1(x')}{\Atrue}{\emptyset}$.
SecRSL introduces the relational assertion~$\Alow{e}{e'}$ to assert that
expressions~$e$ and~$e'$ denote identical values.
We note that~$HP$ and~$HQ$ \emph{under-approximate} the set of $\high$ locations. This explains why the post-set in this judgement is empty.

We defer an explanation of the assertion forms of SecRSL and its rules
to \cref{logic}. However we note that it supports analogues of all of the
reasoning principles of RSL. In particular, atomic locations like~$a$
in \cref{fig:mp} carry (relational) invariants~$Q(v)$ that can be used
to reason about ownership transfer via Release-Acquire patterns.
Being relational, these invariants can express security properties as
well as functional ones.

\begin{figure}
    \begin{center}
    \PQ{$Q(v) \defeq (v=0\land \Aemp)\lor (v=1\land (\Apt{b}{b}{x}{x'} \sep \Alow{x}{x'})); \qquad Q' \defeq Q[1 := \emp]$}\\[1ex]
    \PQ{$[\emptyset], \{\Alow{x}{x'}\}$}\\
    $\Let{b}{\Alloc{\high}}{}$\\
    \PQ{$[b], \{\Alow{x}{x'} \sep \Aptu{b}{b}\}$}\\
    $\Let{a}{\Alloc{\low}}{}$\\
    \PQ{$[b], \{\Alow{x}{x'} \sep \Aptu{b}{b} \sep \Arel{a}{Q} \sep \Aacq{a}{Q}\}$}\\
    $\Let{c}{\Alloc{\low}}{}$\\
    \PQ{$[b], \{\Alow{x}{x'} \sep \Aptu{c}{c} \sep \Aptu{b}{b} \sep \Arel{a}{Q} \sep \Aacq{a}{Q}\}$}\\    
    $\Store{a}{\rlx}{0};$\\
    \PQ{$[b], \{\Alow{x}{x'} \sep \Aptu{c}{c} \sep \Aptu{b}{b} \sep \Arel{a}{Q} \sep \Aacq{a}{Q} \sep \Ainit{a}\}$}\\
    \begin{tabular}{@{}l@{\ }||@{\ }l}
                \PQ{$[b], \{\Alow{x}{x'} \sep \Aptu{b}{b} \sep \Arel{a}{Q} \}$}  & \qquad\PQ{$[b], \{ \Aptu{c}{c} \sep \Aacq{a}{Q} \sep \Ainit{a} \}$} \\                      
                $\qquad \qquad \qquad \Store{b}{\na}{x};$& $\qquad\qquad\qquad\Repeat{\Load{a}{\acq}}{};$\\
                \PQ{$[b], \{\Alow{x}{x'} \sep  \Apt{b}{b}{x}{x'} \sep \Arel{a}{Q} \}$}  & \PQ{$[b], \{\Alow{x}{x'} \sep  \Apt{b}{b}{x}{x'} \sep \Aptu{c}{c} $} \\   
                \ &\PQ{ \qquad \qquad \qquad \qquad$\sep \Aacq{a}{Q'} \}$ }\\                   
                $\qquad \qquad \qquad \Store{a}{\rel}{1};$& $\qquad\qquad\qquad\Let{y}{\Load{b}{\na}}{}$\\
                \qquad \PQ{$[b], \{ \Arel{a}{Q} \sep \Ainit{a} \}$}  & \PQ{$[b], \{\Apt{b}{b}{y}{y} \sep \Aptu{c}{c} \sep \Aacq{a}{Q'} \}$} \\
                & $\qquad\qquad\qquad\qquad\Store{c}{\na}{y}$\\
                & \PQ{$[b], \{\Apt{b}{b}{y}{y} \sep \Apt{c}{c}{y}{y} $} \\
                & \PQ{$\qquad\qquad\qquad\qquad\qquad\qquad\qquad \sep \Aacq{a}{Q'} \}$} \\
    \end{tabular}\\
 \hspace{-1.5em}\PQ{$[\emptyset], \{\Atrue\}$}
\end{center}\par
    \caption{Secure message passing in SecRSL, proving $\Triple{\emptyset}{\Alow{x}{x'}}{\MP_1(x)}{\MP_1(x')}{\Atrue}{\emptyset}$.\label{fig:mp-proof}
    To reduce clutter, we write proof sketches
as if they operate over a single
program. However, they should really be understood
as operating over a pair of programs, where the second program is obtained
by substituting each variable in the first program with its primed
counterpart. This proof sketch uses notational shorthands defined on
page~\pageref{shorthands}.
}
\end{figure}

\cref{fig:mp-proof} depicts a proof sketch for the security of~$\MP_1$
in SecRSL. We note that the invariant~$Q$ used for this proof encodes
that when location~$a$ holds the value 1, then location~$b$ holds a public
value. Hence this example demonstrates SecRSL's ability to support
reasoning about value-dependent classification policies, as well as the local,
compositional reasoning with ownership transfer that makes concurrent
separation logics so powerful.

\section{Language and Memory Model}
\label{C11}

\subsection{Programming Language}
SecRSL  is defined over the Release-Acquire fragment of C11 with relaxed
accesses of RSL~\cite{Vafeiadis_Narayan_13}.
The only difference is that in SecRSL, each memory allocation expression~$\Alloc{F}$
is annotated with a security level $F \in \{\high, \low\}$,
which is used to define the top-level information-flow
security property established by
SecRSL. These annotations are \emph{ghost information} and are ignored
by the language semantics (see \cref{sec:memory-model}).

As with RSL~\cite{Vafeiadis_Narayan_13}, SecRSL programs are assumed to be
in A-Normal form~\cite{flanagan1993essence}. Thus
an atomic expression, $e \in \AExp$, is either a variable~$x$ or a value~$v$
(a memory location~$\ell \in \Loc$ or a number~$n \in \nat$).
Program expressions, $E \in \Exp$, include
atomic expressions~$e$, let-bindings, conditionals, loops, parallel
composition, labelled memory allocations~$\Alloc{F}$, loads~$\Load{e}{X}$, stores~$\Store{e}{Y}{e'}$, and atomic compare-and-swap (CAS) instructions.
\begin{align*}
	v \in \Val ::=\ & \ell\mid n\ \ \ \ \ \ \ \ \ \text{where}~\ell \in \Loc, n\in\nat\\
	e \in \AExp ::=\ & x\mid v\ \ \ \ \ \ \ \ \ \text{where}~x\in \Var\\
	E \in \Exp ::=\ & e \mid \Let{x}{E}{E'} \mid \If{e}{E}{E'}\\
	&\mid \Repeat{E} \mid \Parallel{E_1}{E_2} \mid \Alloc{F}\\
	&\mid \Load{e}{X} \mid \Store{e}{Y}{e'} \mid \CAS{Z}{W}{e}{e'}{e''}\\
	\text{where}~& X\in \{\scx, \acq, \rlx, \na\}, ~Y\in \{\scx, \rel, \rlx, \na\}, \\
	&Z\in \{\scx, \relacq, \acq, \rel, \rlx\},\\
	&W\in \{\scx, \acq, \rlx\}, ~F\in \{\high, \low\}
\end{align*}\par
As in C, in conditional expressions we treat zero as false and non-zero values as true. The construct $\Repeat{E}$ executes $E$ repeatedly until it returns a non-zero value.

In the C11 Release-Acquire memory model,
memory accesses are annotated by their mode, which affects the ordering
guarantees they provide. Modes comprise: sequentially consistent ($\scx$), acquire ($\acq$), release ($\rel$), combined release-acquire ($\relacq$), relaxed ($\rlx$), or non-atomic ($\na$). Different kinds of memory accesses support
different modes. For instance, reads cannot be releases, writes cannot be acquires, CASs cannot be non-atomic.
%

CAS is an atomic operation used heavily in lock-free concurrent algorithms.
In high-level terms, it takes a location, $\ell$, and two values, $v'$ and $v''$, as arguments. It atomically checks if the value in the memory location~$\ell$ is $v'$. If it is then the CAS operation
is said to succeed and it atomically stores $v''$ into memory location $\ell$ and returns~$v'$ (the old value).
Otherwise the CAS is said to fail and it does not modify the memory and simply returns whatever non-$v'$
value was stored in location~$\ell$.  CAS expressions $\CAS{Z}{W}{e}{e'}{e''}$ 
are annotated with two access modes: one~$Z$ to be used for the successful case, and the other~$W$ for the unsuccessful case. The expressions~$e$, $e'$ and~$e''$ respectively denote the values~$\ell$, $v'$ and~$v''$ mentioned above.

We write $\Load{E}{\na}$ to abbreviate
$\Let{x}{E}{\Load{x}{\na}}$; and $E_1; E_2$ for
$\Let{x}{E_1}{E_2}$ when $x$ is not free in~$E_2$.

\subsection{Memory Model}\label{sec:memory-model}
Like other relaxed memory models~\cite{manson2005java,sarkar2009semantics,alglave2009semantics,batty2011mathematizing,mador2012axiomatic,batty2016overhauling}
the semantics that our programming language inherits from RSL~\cite{Vafeiadis_Narayan_13} is defined axiomatically, in which a program's semantics is defined
as a set of its \emph{executions}. Each execution is represented by a set of actions plus various binary relations on those actions.
Actions in our semantics are identical to those in RSL, except
that each allocate action~$\Alab{F}{\ell}$ carries the security label~$F$ of the
allocation expression~$\Alloc{F}$ that generated it.
\begin{center}
	$\Act ::= \skiplab \mid \Wlab{(\scx \mid \rel \mid \rlx \mid \na)}{\ell}{v} \mid \Rlab{(\scx \mid \acq \mid \rlx \mid \na)}{\ell}{v}$\\
	$\mid \RMWlab{(\scx \mid \rel \mid \relacq \mid \acq \mid \rlx)}{\ell}{v}{v'} \mid \Alab{(\low \mid \high)}{\ell}$
\end{center}

\noindent Actions describe a program's interactions with memory. The $\skiplab$ action is a no-op, and represents local computation, as well as thread forks and joins. The other actions represent respectively: writes (generated by store expressions); reads (generated by load expressions); atomic read-modify-write actions (generated by CAS expressions); and allocations. Actions carry information about the memory location~$\ell$ that was accessed and the values~$v,v'$ etc.\ read or written. 

The binary relations over actions include~\cite{Vafeiadis_Narayan_13} the
\emph{sequenced-before relation} $\sbo$, which relates actions according to the order that they appear in the program's textual control flow. We have $\sbo(a, b)$ if~$a$ immediately precedes~$b$. The \emph{reads-from} relation~$\rf$
is a map from read actions~$r$ to write actions~$w$: when $\rf(r) = w$, it means that the value read by the action~$r$ was written to the location that is being read by the action~$w$. The \emph{memory-order}~\cite{Vafeiadis_Narayan_13} relation~$\mo$ (aka the \emph{modification-order}~\cite{Batty:phd} relation) and the
\emph{sequential-consistency} order relation~$\sco$ are total orders: the former
relates write actions to the same atomic location; the latter imposes a total
order on all $\scx$ actions.

The arrows in \cref{fig:mpLab} depict three of these relations:
$\sbo$ (solid, black);
 $\rf$ (dashed, green); and $\mo$ (dotted, orange).

In a program execution, each action is identified by a unique \emph{name},
drawn from some countably infinite set $\AName$. To give meaning to these
opaque action names, the execution includes a 
\emph{labelling} function~$\lab$ that associates
each with an action~$a \in \Act$. Then, formally, an execution is a tuple
$\exec{\A}{\lab}{\sbo}{\rf}{\mo}{\sco}$ where~$\A$ is a finite subset
of action names from $\AName$, $\lab$ is a function from $\AName$ to $\Act$,
$\sbo$ and $\mo$ are binary relations on elements of $\A$,
and~$\rf$ is a partial map between elements of~$\A$.

Two additional relations are defined in terms of the others.
The \emph{synchronises-with} relation~$\sw$:
intuitively $\sw(w,r)$ holds when~$w$ is a release
write and~$r$ is an acquire read that synchronises with~$w$.
We omit its full definition (see~\citet[Figure~3]{Vafeiadis_Narayan_13})
for the sake of brevity.
The \emph{happens-before} relation~$\hb$ formalises when one action must
complete before another, and is the transitive closure of $\sbo \union \sw$.

The semantics of a program~$E \in \Exp$ is defined as usual for axiomatic weak
memory models: the possible executions of~$E$ are calculated and then
constrained by the axioms of the memory model to leave only those
executions that are \emph{consistent} with the model.
We refer to~\citet{Vafeiadis_Narayan_13} for the axioms and
write $\Consistent{\exec{\A}{\lab}{\sbo}{\rf}{\mo}{\sco}}$ when an
execution $\exec{\A}{\lab}{\sbo}{\rf}{\mo}{\sco}$ is consistent.

Executions are calculated by applying a semantic function~$\sem{\cdot}$
that maps expressions~$E \in \Exp$ to a set~$\sem{E}$ of their
executions, each of which is a tuple~$\exec{\res}{\A}{\lab}{\sbo}{\fst}{\lst}$
where~$\A$, $\lab$ and~$\sbo$ are as above, $\res$ is the expression's result,
and~$\fst$ and~$\lst$ respectively denote the (action names of) the first and
last actions of the expression in the~$\sbo$ order.
We also refer readers to~\citet[Section~3]{Vafeiadis_Narayan_13} for the
details of the semantics, which are unchanged except that allocation
expressions~$\Alloc{F}$ produce allocation actions~$\Alab{F}{\ell}$ that
carry their security label~$F$. Note that this allocation label is otherwise
ignored by the semantics, and so does not influence program execution; it is
merely propagated to the label of allocation actions to make it visible to
SecRSL's top-level security property (\cref{adequacy}).

Like RSL, some rules of SecRSL are valid only in a stronger memory model
that strengthens the C11 axioms to exclude so-called
``out of thin air'' reads in programs with relaxed writes.
The definition of this stronger memory model is
identical to that of~\citet[Section 6]{Vafeiadis_Narayan_13}.
We refer to it as the \emph{strengthened} memory model, while noting that
this strengthening is quite standard and was used as the base for various
separation logics that succeeded RSL~\cite{FSL,FSLplusplus}.

\section{Logic}\label{logic}

Like many other logics for proving secure information
flow~\cite{Benton_04,barthe2017proving,Ernst_Murray_19,maillard2019next}, which necessarily
relate pairs of program executions, SecRSL is a \emph{relational} logic~\cite{Benton_04}. In general, SecRSL reasons about a pair of A-Normal form
programs~$E$ and~$E'$ to prove that they are indistinguishable 
to the
attacker and thus do not leak sensitive information.
(We define indistinguishability in \cref{adequacy}.)

\newcommand{\hi}{\mathit{hi}}
\newcommand{\lo}{\mathit{lo}}
For example, consider some program~$\Prog_1(\hi,\lo)$ whose initial
state comprises some secret $\hi$, as well as some public data~$\lo$.
Security requires that when run from two initial states that agree on
$\lo$ but might differ on~$\hi$, the executions of $\Prog_1$ are
indistinguishable.
Therefore,
to prove this hypothetical program secure, we must prove that for all
values~$\lo$, $\hi$, $\lo'$, $\hi'$, if $\lo$ and~$\lo'$ are equal then
the behaviours of $\Prog_1(\lo,\hi)$ are indistinguishable to the attacker
from the behaviours of $\Prog_1(\lo',\hi')$.

\paragraph{SecRSL Judgement}
Thus the SecRSL judgement:
\[
\Triple{HP}{P}{E}{E'}{(y,y').\ Q}{(y,y').\ HQ}
\]
where $E$ and~$E'$ are
expressions (programs). Intuitively, this judgement means that from the
attacker's point of view, the executions
of~$E$ are indistinguishable from those of~$E'$. $P$ (respectively $Q$) is
a \emph{relational} separation logic precondition~\cite{Ernst_Murray_19}
(respectively postcondition, which may refer to the values~$y$ and~$y'$
returned by the expressions~$E$ and~$E'$ respectively).
We introduce SecRSL's assertion
language shortly. $HP$ (respectively $HQ$) under-approximates the set of
memory locations known to be~$\high$ (i.e.\ unobservable to the attacker)
before (respectively after)~$E$ and~$E'$
execute. Like~$Q$, $HQ$ can also refer to the return values~$y$ and~$y'$ of
$E$ and~$E'$ respectively.\footnote{That SecRSL tracks which locations
  are known to be unobservable to the attacker, rather than tracking those
  that are known to be observable, might be surprising. We explain this
  design choice later in \cref{sec:discussion}.} 
As with prior security logics (e.g. \SecCSL~\cite{Ernst_Murray_19}), the
sets $HP$ and~$HQ$ can depend on values read during the program and so
support value-dependent location sensitivity~\cite{Murray_SPR_16}
(i.e.\ locations whose classification depends on runtime values).

\paragraph{Notational Shorthands}\label{shorthands}
For boolean condition~$B$ and SecRSL assertion~$P$, we write $P\land B$ to abbreviate
$\metaITE{B}{P}{\Afalse}$.
When writing postconditions we will write
$\{(y,y).\ Q\}$ instead of $\{(y,y').\ Q\land y=y'\}$, or omit
the ``$(y,y').$'' when the postcondition does not refer to the return
values. To save space, we also avoid duplicating the~``$(y,y').$'' prefix
in both the postcondition and the post-set of $\high$ locations, omitting it
from the latter. We write $\TripleNA{P}{E}{E'}{(y,y').\ Q}$,
when~$E$ and~$E'$
neither modify nor depend on the $\high$ locations set, which
abbreviates
$\forall HP.\ \Triple{HP}{P}{E}{E'\linebreak}{(y,y').\ Q}{HP}$.
We write $\TripleAH{HP}{P}{E}{(y,y').\ Q}{HQ}$, when talking about two
identical programs~$E$, to abbreviate 
$\Triple{HP}{P}{E}{E}{(y,y').\ Q}{HQ}$. We also combine these abbreviations:
$\TripleA{P}{E}{(y,y').\ Q}$ means $\forall HP.\ \Triple{HP}{P}{E}{E}{(y,y').\ Q}{HP}$.
These conventions are used both for presenting the rules of the logic and
to simplify proof sketches like that of~\cref{fig:mp-proof}.

\paragraph{SecRSL Assertions}
SecRSL's relational assertions~$P, Q$ etc. are evaluated over pairs of memories, like
\SecCSL's assertions~\cite{Ernst_Murray_19}. They include relational (i.e. 2-state) analogues
of all the RSL assertions~\cite{Vafeiadis_Narayan_13} and are defined below.
\begin{center}
	$P,P_1,P_2 ::= \Afalse \mid \Aimplies{P_1}{P_2} \mid \Aforall{x}{x'}{P} \mid \Apt{\ell}{\ell'}{e}{e'}$\\
	$\mid \Aemp \mid \Astar{P_1}{P_2} \mid \Ainit{\ell} \mid \Aptu{\ell}{\ell'} \mid \Alow{e}{e'}$\\
	$\ \mid \Arel{\ell}{Q} \mid \Aacq{\ell}{Q} \mid \Armw{\ell}{Q}$
\end{center}

In a judgement $\Triple{HP}{P}{E}{E'}{(y,y').\ Q}{(y,y').\ HQ}$ we call
$E$ the \emph{left} program and~$E'$ the \emph{right} program. SecRSL assertions
are therefore evaluated in pairs of heaps (memories):
the left (respectively right)
memories are those encountered during the execution of~$E$ (respectively $E'$).
Thus universal quantification quantifies over both a left and a right variable;
points-to assertions $\Apt{\ell}{\ell'}{e}{e'}$ state that in the left memory,
the non-atomic location~$\ell$ holds the value denoted by~$e$ (and likewise for the right
memory with~$\ell'$ and~$e'$). Likewise $\Aptu{\ell}{\ell'}$ means that
the non-atomic location~$\ell$ ($\ell'$) is uninitialised in the left (right) memory. $\Alow{e}{e'}$ means that $e$ and~$e'$ denote equal values, and so
are indistinguishable to the attacker and, hence, can be treated as public
data that is safe to reveal to the attacker without violating security. It is
analogous to \SecCSL's \emph{value-sensitivity} assertion~\cite{Ernst_Murray_19}.

As in RSL, the assertions $\Ainit{\ell}$,
$\Arel{\ell}{Q}$, $\Aacq{\ell}{Q}$ and $\Armw{\ell}{Q}$ refer to atomic locations~$\ell$.
$\Ainit{\ell}$ means that atomic location~$\ell$ has been initialised; the
others
represent the permission to perform the corresponding action on that
location.
$Q$ is an invariant, attached to atomic location~$\ell$, and is
parameterised by the value~$v$ in memory at location~$\ell$.
For instance, $\Aacq{\ell}{Q}$ is the permission
to perform an acquire load on location~$\ell$, obtaining the invariant~$Q(v)$
where~$v$ is the value read from the location; $\Arel{\ell}{Q}$ is the
permission to perform a release write to location~$\ell$, provided that the
invariant~$Q(v)$ holds where~$v$ is the value being written; the $\Armw{\ell}{Q}$
permission allows performing the CAS operation on location~$\ell$. 

Unlike the aforementioned assertions for non-atomic locations, those for
atomic locations do not need to mention pairs of locations~$(\ell,\ell')$
or pairs of values~$(v,v')$. This is because SecRSL requires that each atomic
location is $\low$ and that the access patterns of which locations are
accessed are identical between the left and right executions
(\cref{sec:overview}). Thus atomic locations will be proved to be
identical in the left and right
executions (both the location~$\ell$ itself and the value~$v$ that it holds).

From these assertions the standard logical connectives
$\Atrue$, $\land$, $\lor$, $\lnot$, and $\exists$ can be derived
in the usual way.
Moreover, SecRSL assertions enjoy the
usual semantic equivalences of separation logic. In particular, 
we have that $\forall P,Q,R\in \Assn:$\\[-1.2em]
\begin{align*}
	P\sep Q\sim&\ Q\sep P\\
	P\sep (Q\sep R)\sim&\ (P\sep Q) \sep R\\
	P\sep \Aemp\sim&\ P\\
	\Afalse \sep \Afalse \sim&\ \Afalse\\
	P\lor Q\sim&\ Q\lor P\\
	P\lor (Q\lor R)\sim&\ (P\lor Q) \lor R\\
	(P\lor \Afalse)\sim&\ (P\lor P)\\
	(P\lor P)\sim&\ P
\end{align*}
These equivalences are derived from the assertion 
semantics, which will be introduced in \cref{assn_valid} \ifExtended (see also \cref{app:assertion})\fi.

\begin{figure*}[!htp]
	\begin{mathpar}
		\infer[\textsc{A-L}]
		{\begin{array}{c}
				\TripleA{\Aemp}{\Alloc{\low}}{(\ell,\ell).\Aptu{\ell}{\ell}}
		\end{array}}
		{}
		
		\infer[\textsc{A-H}]
		{\begin{array}{c}
				\TripleAH{HF}{\Aemp}{\Alloc{\high}}{(\ell,\ell).\Aptu{\ell}{\ell}}{\{\ell\}\union HF}
			\end{array}
		}
		{}
		
		\infer[\textsc{NA-W}]
		{\begin{array}{c}
			\Triple{HF}{P \sep G}{\Store{\ell}{\na}{v}}{\Store{\ell}{\na}{v'}}{\Apt{\ell}{\ell}{v}{v'}}{HF}
		\end{array}
		}
		{\begin{array}{c}
				\metaITE{\ell\not\in HF}{G = \Alow{v}{v'}}{G = \Aemp} \\
				(P=\Aptu{\ell}{\ell})\  \lor \  (P=\Apt{\ell}{\ell}{\_}{\_})
		\end{array}}
		
		\infer[\textsc{NA-R}]
		{\begin{array}{c}
			\TripleNA{\Apt{\ell}{\ell}{v}{v'}}{\Load{\ell}{\na}}{\Load{\ell}{\na}}{(y,y'). \Apt{\ell}{\ell}{v}{v'}\land y=v\land y'=v'}
		\end{array}
		}
		{}
		
		\infer[\textsc{Value}]
		{\TripleNA{P}{e}{e'}{(y,y').P\land y = e \land y' = e')}}
		{}
	\end{mathpar}
	\caption{Non-Atomic Rules}
	\label{NaRules}
\end{figure*}

\begin{figure*}[!htp]
\begin{DIFnomarkup}
	\begin{mathpar}
		\infer[\textsc{A-R}]
		{\TripleA{\Aemp}{\Alloc{\low}} {(\ell,\ell).\Arel{\ell}{Q} \sep \Aacq{\ell}{Q}}}
		{}
		
		\infer[\textsc{A-M}]
		{\TripleA{\Aemp}{\Alloc{\low}} {(\ell,\ell).\Arel{\ell}{Q} \sep \Armw{\ell}{Q}}}
		{}
		
		\infer[\textsc{Rlx-R}]
		{\TripleA{\Aacq{\ell}{Q}\sep \Ainit{\ell}}{\Load{\ell}{\rlx}} {(v,v). \Aacq{\ell}{Q}}}
		{}
		
		\infer[\textsc{Rel-W}]
		{\TripleA{\Arel{\ell}{Q}\sep Q(v)}{\Store{\ell}{\rel}{v}} {\Arel{\ell}{Q}\sep \Ainit{\ell}}}
		{}
		
		\infer[\textsc{Acq-R}]
		{\TripleA{\Aacq{\ell}{Q}\sep \Ainit{\ell}}{\Load{\ell}{\acq}}{(v,v).\ Q(v) \sep \Aacq{\ell}{Q\left[v:=\Aemp\right]}}}
		{\forall v.~\Precise{Q(v)}}
		
		\infer[\textsc{Rlx-W*}]
		{\hoare{\Arel{\ell}{Q}}{[\ell]_{\rlx}:=v} {\text{Init}(\ell)}}
		{Q =\lambda x.~(\metaITE{x=v}{\emp}{\False})}
		
		\infer[\textsc{Rlx-R*}]
		{\TripleA{\Aacq{\ell}{Q}\sep \Ainit{\ell}}{\Load{\ell}{\rlx}}{(v,v).~\Aacq{\ell}{Q}\land Q(v)\neq\Afalse }}
		{}
		
		\infer[\textsc{CAS*}]
		{\TripleA{P}{\CAS{X}{Y}{\ell}{v}{v'}} {(y,y).~R}}
		{\begin{array}{c}
				P \implies \Ainit{\ell} \sep \Armw{\ell}{Q} \sep \Atrue \\
				P \sep Q(v) \implies \Arel{\ell}{Q'} \sep Q'(v') \sep R[v/y] \\
				X\in \{\rel,\rlx\} \implies Q(v) = \Aemp \\
				X\in \{\acq,\rlx\} \implies Q'(v') = \Aemp \\
				\TripleA{P}{\Load{\ell}{Y}}{(y,y).~ y\neq v \Rightarrow R}
		\end{array}}
		
		\infer[\textsc{RELAX}]
		{\Triple{HP}{P}{E_T}{E_T}{(y,y).Q}{HQ}}
		{\Triple{HP}{P}{E_t}{E_t}{(y,y).Q}{HQ}\ \ \ \ \ \ \ t\sqsubseteq T}\ , \text{where}\ \rlx\sqsubseteq\rel\sqsubseteq\scx \text{ and } \rlx\sqsubseteq\acq\sqsubseteq\scx
		
	\end{mathpar}
	\caption{Atomic Rules}
	\label{ARules}
\end{DIFnomarkup}
\end{figure*}

\paragraph{Non-Atomic Rules}
The rules for operations on non-atomic locations are shown in
\cref{NaRules}.
Recall that our threat model assumes a memory allocator
that does not leak sensitive information, i.e. the resolution of the nondeterministic
choice about which memory location the~$\Alloc{}$ expression returns does
not reveal secrets. Additionally, recall from \cref{sec:overview}
that SecRSL does not allow
programs to branch on secrets. This ensures that
all decisions about whether to
allocate or not depend only on public data. Hence, when reasoning about
memory allocation (rules \textsc{A-L} and \textsc{A-H}), the postcondition can assert that the two locations~$\ell$
and~$\ell'$ returned in the left and right executions respectively are equal.
Naturally, $\high$ allocations increase the set of known $\high$ locations
(see rule~\textsc{A-H}).

When writing to a non-atomic location~$\ell$,
the rule \textsc{NA-W} ensures
security by requiring that if the write might be visible to the attacker
($\ell \not\in HF$)
then
the values~$v$ and~$v'$ being written must be indistinguishable: $\Alow{v}{v'}$.

In addition notice that the rules for reading and writing non-atomic
locations require that the location~$\ell$ being accessed is identical
in both programs. This is required to provide SecRSL's constant-time
guarantee of indistinguishability against attackers who can observe
the memory access pattern of the program (\cref{sec:overview}).

The \textsc{Value} rule in \cref{NaRules}
is analogous to its RSL counterpart. 

\begin{figure}[!htp]\[
	\begin{array}{c}
		\Aemp\iff \Alow{x}{x}\\[1ex]
		\Ainit{\ell}\iff \Ainit{\ell}\sep\Ainit{\ell}\\[1ex]
		\Arel{\ell}{\lambda v.~Q_1(v)\lor Q_2(v)}\iff \Arel{\ell}{Q_1} \sep \Arel{\ell}{Q_2}\\[1ex]
		\Aacq{\ell}{\lambda v.~Q_1(v)\sep Q_2(v)}\iff \Aacq{\ell}{Q_1} \sep \Aacq{\ell}{Q_2}\\[1ex]
		\Armw{\ell}{Q}\iff \Armw{\ell}{Q} \sep \Armw{\ell}{Q}\\
		\\
		\infer[]
		{\Armw{\ell}{Q}\iff \Armw{\ell}{Q} \sep \Aacq{\ell}{Q'}}
		{\forall v.~ (Q'(v) = \Aemp \lor Q(v) = Q'(v) = \Afalse)}
	\end{array}\]
	\caption{Split Rules}
	\label{ASRules}
\end{figure}

\begin{figure*}[!htp]
	\begin{mathpar}
		
		\infer[\textsc{Conseq}]
		{{\Triple{HP}{P}{E}{E'}{(y,y').Q}{HQ}}}
		{\begin{array}{c}
				(P  \implies P') ~\land~ HP'\subseteq HP \\
				~\forall (y,y').~~(Q' \implies Q) ~\land~HQ\subseteq HQ'\\
				\Triple{HP'}{P'}{E}{E'}{(y,y').Q'}{HQ'}
			\end{array}
		}
		
		\infer[\textsc{DISJ}]
		{\Triple{HP}{P\lor P'}{E}{E'}{(y,y').Q\lor Q'}{HQ}}
		{\begin{array}{c}
				\Triple{HP}{P}{E}{E'}{(y,y').Q}{HQ} \\
				\Triple{HP}{P'}{E}{E'}{(y,y').Q'}{HQ}\\
		\end{array}}
		
		\hspace*{-1.5em}\infer[\textsc{If}]
		{\Triple{\IF\ b\ \THEN\ H1\ \ELSE\ H2}{\Alow{b}{b'} \sep P}{E}{E'}{(y,y').~Q}{HQ}}
		{\begin{array}{c}
				\Triple{H1}{b \land b' \land P}{E_1}{E_1'}{(y,y').Q}{HQ}\\
				\Triple{H2}{\lnot b \land\lnot b' \land P}{E_2}{E_2'}{(y,y').Q}{HQ}\\
				E=\IF\ b\ \THEN\ E_1\ \ELSE\ E_2\qquad\qquad E'= \IF\ b'\ \THEN\ E_1'\ \ELSE\ E_2'
		\end{array}}
		
		\infer[\textsc{EX}]
		{\Triple{HP}{\exists x.P}{E}{E'}{(y,y').\exists x.Q}{HQ}}
		{\Triple{HP}{P}{E}{E'}{(y,y').Q}{HQ}}
		
		\infer[\textsc{FRAME}]
		{\Triple{HP}{P \sep F}{E}{E'}{(y,y').Q \sep F}{HQ}}
		{\Triple{HP}{P}{E}{E'}{(y,y').Q}{HQ}}
		
		\infer[\textsc{Par}]
		{\Triple{HP}{P_1 \sep P_2}{\Parallel{E_1}{E_2}}{\Parallel{E_1'}{E_2'}}{(y,y').Q_1 \sep Q_2}{HQ_1\union HQ_2}}
		{\begin{array}{c}
				\Triple{HP}{P_1}{E_1}{E_1'}{(y,y').Q_1}{HQ_1} \\
				\Triple{HP}{P_2}{E_2}{E_2'}{Q_2}{HQ_2}
		\end{array}}
		
		\infer[\textsc{REPEAT}]
		{\TripleNA{P}{\Repeat{E}}{\Repeat{E'}}{(y,y).Q \land y\neq 0}}
		{\begin{array}{c}
				\TripleNA{P}{E}{E'}{(y,y').~Q \sep \Alow{y}{y'}}\\
				Q[(0,0)/(y,y')]  \implies P
		\end{array}}
		
		\infer[\textsc{LET}]
		{\Triple{HP_1}{P_1}{\Let{x}{E_1}{E_2}}{\Let{x'}{E_1'}{E_2'}}{(y,y').Q_2}{HQ_2}}
		{\begin{array}{c}
				\Triple{HP_1}{P_1}{E_1}{E_1'}{(x,x').Q_1}{HQ_1}\\
				\forall~(x ,x').\Triple{HQ_1}{Q_1}{E_2}{E_2'}{(y,y').Q_2}{HQ_2}
		\end{array}}

	\end{mathpar}
	\caption{Structural Proof Rules}
	\label{STRules}
\end{figure*}

\paragraph{Atomic Rules}
The rules for operations on atomic memory locations are depicted in
\cref{ARules}, and are each analogous to their RSL
counterparts. Rules for allocating atomic locations (\textsc{A-R} and \textsc{A-M}), like their non-atomic
counterparts, assert that the chosen location is identical (and thus public)
in both executions, returning the appropriate permissions as in RSL.
Notice that these rules are valid only for $\low$ allocations, since SecRSL
requires that all atomic locations are~$\low$ (\cref{sec:overview}). That is also the reason why
each of these rules talks about
two identical programs, since SecRSL requires that all actions on $\low$
locations are identical between the two executions.
As in RSL, performing a release write (rule \textsc{Rel-W}) requires the
permission~$\Arel{\ell}{Q}$ as well as evidence that the invariant~$Q(v)$
holds for the value~$v$ being written, which must be public (equal in both
executions), as must the location~$\ell$ being written to. Similarly,
performing an acquire read (rule \textsc{Acq-R}) requires the appropriate
permission plus evidence that the location was initialised. Doing so obtains
the invariant~$Q(v)$ for the value~$v$ that was read, which is identical in
both executions (because atomic locations must be~$\low$). As in RSL this
rule requires that~$Q$ is precise, whose definition for SecRSL's relational
assertions we postpone to \cref{defn:precise} in \cref{assn_valid}.
The rules for relaxed
reads and writes also follow their RSL counterparts. As there, the rules
\textsc{RLX-R*}, \textsc{RLX-W*}, and \textsc{CAS*} (when $X\notin\{\relacq,\scx \}$) are only sound in the \emph{strengthened} memory model (\cref{sec:memory-model}) of RSL and its successors~\cite{FSL,FSLplusplus}. 

\paragraph{Split Rules}
SecRSL assertions behave identically to their RSL counterparts. The
assertions on atomic locations behave identically as in RSL
(see \cref{ASRules}): $\Ainit{\ell}$ and $\Armw{\ell}{Q}$ are duplicable, as is $\Arel{\ell}{Q}$ which can also
be split along disjunctions in~$Q$; $\Aacq{\ell}{Q}$ on the other hand can be split along
separating conjunction in~$Q$. $\Armw{\ell}{Q}$ can also be split to produce a
separate, but weak, $\Aacq{\ell}{Q}$ permission.

\paragraph{Structural Rules}
The structural rules for SecRSL are shown in \cref{STRules}. It
supports all the standard rules from concurrent separation logic, like RSL
before it. Notice that the consequence rule \textsc{Conseq} allows
one to expand the pre-set $HP$ of $\high$ locations, or reduce the post-set~$HQ$.  As in \SecCSL~\cite{Ernst_Murray_19}, the rules for conditionals and
loops do not allow programs to branch on secrets. This is a necessary
condition to enforce SecRSL's constant-time security (\cref{sec:overview}).

As with RSL, the rule for parallel composition~\textsc{Par} is not symmetric.
This is because we inherit RSL's semantics for parallel composition in which,
for $\Parallel{E_1}{E_2}$, the return value of $E_2$ is discarded. This is why
the postcondition~$Q_2$ in the second premise of the \textsc{Par} rule does
not refer to the return values of~$E_2$ and~$E_2'$.

\section{Soundness}
\label{soundness}

Like all program logics, SecRSL's soundness is proved against the programming
language semantics. Specifically, as with prior security separation logics
like \SecCSL~\cite{Ernst_Murray_19}, we define a set of top-level security
and safety
properties over the programming language semantics that apply to entire
program executions. We prove that if a
program is verified in SecRSL then it enjoys these properties.
We call this result SecRSL's \emph{adequacy theorem} (\cref{thm:adequacy}),  in \cref{adequacy}. 

In order to prove these properties, we developed a compositional
and inductive
definition of relational validity over the C11 weak memory model
that encodes the semantic
meaning of SecRSL judgements~$\Triple{HP}{P}{E}{E'}{(y,y').\ Q}{HQ}$.
We call this property \emph{relational validity}.
It holds when for every consistent execution (i.e.\ one that is consistent
with the axioms of the memory model)
of~$E$, one can \emph{inductively} construct a related consistent
execution of~$E'$ that witnesses
the security of both executions (by guaranteeing
that they are indistinguishable) and their safety. 
The rules of the logic are then proved sound
against this definition of relational validity.

Naturally, this relational validity rests on the semantics
of SecRSL's assertions, which we present in \cref{assn_valid}.
We present the definition of semantic validity for SecRSL's judgements
in \cref{triple}.

As mentioned, SecRSL's soundness proof was mechanised in Coq, and is available
as supplementary material.

\subsection{Adequacy}
\label{adequacy}

SecRSL's adequacy theorem
includes SecRSL's top-level security guarantees of \emph{low safety}
and \emph{location safety}: the former guarantees that an attacker who can
observe the contents of $\low$ memory locations throughout a program's
execution cannot learn sensitive information. The latter strengthens this
guarantee to also include attackers who can observe the memory-access pattern
of the program, i.e.\ can observe which locations are being accessed but not
the values being written to them. This latter property provides protection
against attackers who e.g.~can mount cache timing channel attacks, and so
provides a form of \emph{constant-time} security~\cite{barthe2019formal}.

To our knowledge, these are the first top-level confidentiality
properties defined over an axiomatic weak memory semantics. We believe that our
properties are of independent interest and could be adapted to 
other axiomatic weak memory models.

The top-level properties that SecRSL guarantees via its adequacy theorem
also include that programs
are memory safe, never read from uninitialised locations,
and are free of data races. These properties SecRSL inherits from
RSL~\cite{Vafeiadis_Narayan_13}.

\paragraph{Whole-program executions}
SecRSL's top-level properties are defined over whole-program executions.
For a program~$E$, we denote this set of executions $\CCof{E}$ 
(the set of \emph{complete consistent} executions~\cite{Vafeiadis_Narayan_13}).
As in RSL, whole-program executions of~$E$
are obtained from the consistent executions of~$E$ by adding two extra
$\skiplab$ actions: one that precedes the execution of~$E$ and another that
follows it.

\begin{definition}[Whole-program executions]\label{ccof}
	We denote by $\CCof{E}$ the set of whole-program executions of~$E$:\\
	$\CCof{E} \defeq \{ \rexec{\res}{\A}{\lab}{\sbo}{\rf}{\mo}{\sco} \mid $\\
	\hspace*{2.2cm} $\exists a,b,\A_{prg},\lab_{prg},\sbo_{prg},\fst,\lst. $
	 $a\neq b \land \A = \{a,b\}\uplus \A_{prg}\ \land$\\
	\hspace*{2.2cm} $\sbo = \sbo_{prg}\union (a,\fst)\union (\lst,b)\ \land$
	$\lab = \{\functo{a}{\skiplab},\functo{b}{\skiplab}\} \union \lab_{prg}\ \land$\\
	\hspace*{2.2cm} $\exec{\res}{\A_{prg}}{\lab_{prg}}{\sbo_{prg}}{\fst}{\lst} \in \setof{E}\ \land$
	$\Consistent{\exec{\A}{\lab}{\sbo}{\rf}{\mo}{\sco}}\}$\\
\end{definition}

\paragraph{Top-level Security Statement}

A key challenge in phrasing the top-level adequacy statement over the C11
memory model, which would persist also for other axiomatic memory models,
is dealing with the nondeterminism inherent in such models. Put simply,
for two programs~$E$ and~$E'$ that (due to nondeterminism) each have a
set of behaviours as their semantics, what does it mean for the behaviours
of~$E$ to be indistinguishable from those of~$E'$?

Following prior work on security definitions for concurrent, nondeterministic
programs~\cite{Murray_Lowe_09} one way to define this is to assume that sources
of nondeterminism will be resolved in the same way in the two programs, and so
to compare only those pairs of executions that arise from resolving the
nondeterminism identically in both programs. $E$ and~$E'$ are indistinguishable
when each such pair of executions is indistinguishable.
If we take this approach for SecRSL, what
assumptions would it hard-wire into
the resulting adequacy theorem?

The sources of nondeterminism in the C11 memory model of
\cref{sec:memory-model} that we inherit from~\citet{Vafeiadis_Narayan_13}
are the nondeterministic model of memory allocation and the weak memory
concurrency semantics. The former abstracts away from how memory locations
are allocated in the $\Alloc{}$ expression. The latter abstracts away from
the concurrency inherent in the C11 language and, hence, the implementation
choices of the compiler that generates code for the underlying instruction
set architecture and its weak memory model, as well as the internal details
of how threads are scheduled by the operating system and hardware.

Considering only pairs of executions in which nondeterminism is resolved
identically therefore encodes into the security property the basic assumption
of \cref{sec:overview} that the memory allocator
never leaks sensitive information, and nor does the
implementation choices of the weak memory model.

What does it mean for nondeterminism to have been resolved identically
in two executions?
To answer this question, let us rephrase what it means to prove
indistinguishability of each pair of executions of~$E$ and~$E'$ in which
nondeterminism has been resolved identically. This assumption can be stated
equivalently by saying that if $E'$ has the opportunity to make an identical
choice to~$E$, then it will do so (and vice-versa). Thus it suffices to
show for each execution of~$E$, there exists a corresponding execution
of~$E'$ in which it is evident that the nondeterministic choices have been
resolved identically, and then prove that these two executions are
indistinguishable.

This is precisely what SecRSL's adequacy theorem does (as well as proving
that both executions are safe).

To do so, we must define a sufficient condition that allows us to judge
when the nondeterminism in two executions has been resolved identically.
Recall that in our semantics (\cref{sec:memory-model}), each execution
is a set of actions related by various binary relations. We can view
each execution therefore as a directed graph (see e.g.~\cref{fig:mpLab}),
whose nodes are the actions
and whose edges carry labels like~$\sbo$, $\rf$, etc.

We claim that nondeterminism has been resolved identically in two
executions when (1)~those executions are \emph{isomorphic} (i.e. there is
one-to-one mapping between the nodes of the two graphs and the two
graphs have the same structure and edge relationships) and (2)~if for
every allocation action~$\Alab{F}{\ell}$ in one graph, the isomorphic
action of the other graph is identical (i.e. is also $\Alab{F}{\ell}$).
Condition~(1) implies identical resolution of nondeterminism resulting
from the weak memory concurrency model, while (2) implies identical
allocation decisions.\pengbo{At here my understanding is that (1) and (2) are both assumptions, but --- see next comment}

Thus SecRSL's adequacy theorem says that for every execution of~$E$, there
exists an execution of~$E'$ for which conditions~(1) and~(2) hold,
and that the
two executions are indistinguishable.

Since our definition of indistinguishability implies condition~(2),
SecRSL's adequacy theorem is stated more succinctly by saying that
for each whole-program execution $\rchi \in \CCof{E}$, there exists an isomorphic whole-program
execution $\rchi' \in \CCof{E'}$ that is indistinguishable to~$\rchi$.

Two executions are isomorphic when they share the same set of
nodes and have identical edge relationships, including for the
derived relations~$\hb$ and~$\sw$. In the graph analogy,
the set of nodes corresponds to the set of opaque action names~$\A$.
Hence we have:

\begin{definition}[Isomorphic]
	Two whole-program executions $\rchi = \rexec{\res}{\A}{\lab}{\sbo}{\rf}{\mo}{\sco}$ and
	$\rchi' = \rexec{\res'}{\A}{\lab'}{\sbo}{\rf}{\mo}{\sco}$ are \emph{isomorphic} if
	and only if for all $a,b \in \A$,\\
	$(\hb_{\rchi}(a,b) \iff \hb_{\rchi'}(a,b)) \land (\sw_{\rchi}(a,b) \iff \sw_{\rchi'}(a,b))$.
\end{definition}

Execution indistinguishability is defined by the two aforementioned properties
of location safety and low safety. The former implies \pengbo{---but at here we say our adequacy implies those assumption, which I don't understand}condition~(1) above
(identical allocation decisions) and implies that the executions are
indistinguishable to an attacker who can observe which locations are
being accessed but not the contents of those locations. It does so by
saying that the labels on isomorphic actions must either match or, if not,
they must both be a non-atomic write (respectively read) to the same location.

\begin{definition}[Location Safety]\label{defn:locsafe}
	Two isomorphic whole-program executions $\rchi = \rexec{\res}{\A}{\lab}{\sbo}{\rf}{\linebreak\mo}{\sco}$ and
	$\rchi' = \rexec{\res'}{\A}{\lab'}{\sbo}{\rf}{\mo}{\sco}$ satisfy
	\emph{location safety} if
	and only if\\
	for all $a\in \A.\ \lab(a) = lab'(a)\ \lor$ \\
	$\qquad (\exists\  \ell\  v\  v'.\ \lab(a) = \Rlab{\na}{\ell}{v}\land \lab'(a) = \Rlab{\na}{\ell}{v'})\ \lor $\\
	\qquad  $(\exists\  \ell\  v\  v'.\ \lab(a) = \Wlab{\na}{\ell}{v}\land \lab'(a) = \Wlab{\na}{\ell}{v'}) $
\end{definition}

The second indistinguishability property \emph{low safety} defines
indistinguishability against an attacker who can observe $\low$ memory
locations. It says that any $\low$ location must be accessed identically
(including the operations performed on that location and the values
written or read to it) between the two executions.

\begin{definition}[Low Safety]\label{defn:lowsafe}
	Two isomorphic whole-program executions $\rchi = \rexec{\res}{\A}{\lab}{\sbo}{\rf}{\linebreak\mo}{\sco}$ and
	$\rchi' = \rexec{\res'}{\A}{\lab'}{\sbo}{\rf}{\mo}{\sco}$ satisfy
	\emph{low safety} if
	and only if for all locations~$\ell$ for which
	$(\exists a\in \A.\ \lab(a) = \Alab{\low}{\ell} \lor \lab'(a) = \Alab{\low}{\ell})$ it is the case that\\
	for all $b\in \A$ such that $\lab(b)$ or $\lab'(b)$ is an action on $\ell$, then $\lab(b) = \lab'(b)$.
\end{definition}

With these definitions, SecRSL's adequacy theorem can be formally stated.
\begin{theorem}[Adequacy]\label{thm:adequacy}
	Let $E$ and $E'$ be two programs such that, for some~$Q$ and~$HQ$, if a SecRSL judgement
	$\Triple{\emptyset}{\Atrue}{E}{E'}{(y,y'). Q}{(y,y'). HQ}$ holds,
	then, for every whole-program execution $\rchi \in \CCof{E}$ there exists some
	$\rchi' \in \CCof{E'}$ such that $\rchi$ and~$\rchi'$ are isomorphic,
	and satisfy location safety and low safety. Moreover, each execution
	is memory safe, has no reads from uninitialised locations, and no data races.
\end{theorem}

We omit the definitions of the safety properties (namely
memory safety, plus absence
of uninitialised reads and data races), since they are
unchanged from RSL~\cite[Section~7.3]{Vafeiadis_Narayan_13}.

\subsection{Semantics of Assertions and Local Relational Validity}
\label{assn_valid}

As in RSL and its descendants~\cite{FSL,FSLplusplus}, SecRSL assertions
are given a semantics over heaps (memories)~$h$. Because the axiomatic
weak memory model
defines executions only in terms of actions and relations
between those actions, memories do not appear in the semantics. As with
logics like RSL, we define what it means for a pair of executions~$\rchi$, $\rchi'$ to
be valid with respect to a judgement $\Triple{HP}{P}{E}{E'}{(y,y).\ Q}{HQ}$
by asserting the existence of memories that annotate the edges of the
executions, and are consistent with the actions being performed, and satisfy
the SecRSL assertions in the judgement. 

We reuse the heap model of RSL~\cite[Section~7.1]{Vafeiadis_Narayan_13}. However, since SecRSL assertions
are relational, each assertion is evaluated over a \emph{pair}~$(h,h')$ of
heaps~\cite{Ernst_Murray_19,yang2007relational}.
Thus the meaning of an assertion~$P$ we denote $\setof{P}$ and define
$\setof{-}$ inductively as a function that maps an assertion to a set of
pairs~$(h,h')$ for which the assertion holds.

The full definition appears in
\ifExtended \cref{app} as \cref{defn:assn-sem}\fi. For
instance, the
meaning of the points-to assertion~$\Apt{\ell}{\ell'}{v}{v'}$
for non-atomic locations is analogous
to its \SecCSL counterpart and holds for a pair of heaps~$(h,h')$ when
$\ell \mapsto v$ in~$h$ and $\ell' \mapsto v'$ in~$h'$:
\[
\setof{\Apt{\ell}{\ell'}{v}{v'}} \defeq\ \{(\{\mapsNA{\ell}{v}\},\{\mapsNA{\ell'}{v'}\})\}
\]

The assertion~$\Alow{e}{e'}$ requires that~$e$ and~$e'$ are indistinguishable
to the attacker and is equivalent to $\metaITE{e=e'}{\emp}{\Afalse}$:
\[
\setof{\Alow{e}{e'}} \defeq\ \metaITE{e=e'}{\{(\emptyset,\emptyset)\}}{\emptyset}
\]
This assertion we include as SecRSL's analogue of \SecCSL's \emph{value sensitivity}
assertion~$\cdot :: \texttt{low}$.

From the relational assertion semantics, we lift the traditional notion
of what it means for a separation logic assertion to be \emph{precise}~\cite{OHearn_04,Vafeiadis_Narayan_13} to
SecRSL's relational assertions.

\begin{definition}[Precise]\label{defn:precise}
	A SecRSL assertion~$P$ is \emph{precise}, written $\Precise{P}$, if and only if\\
	For all $h_1, h_2$ such that $((h_1,\_)\in\setof{P}\land (h_2,\_)\in\setof{P})$ or $((\_,h_1)\in\setof{P}\land (\_,h_2)\in\setof{P})$,\\
    for all $h_3, h_4$ such that $h_1\oplus h_3 = h_2\oplus h_4 \neq \undefine$, we have $h_1 = h_2\land h_3 = h_4$\\
\end{definition}

In this definition~$\oplus$ refers to the heap addition operator of RSL's
heap model~\cite[Figure~15]{Vafeiadis_Narayan_13}.

\paragraph{Local Relational Validity}
The concept of \emph{local relational validity} is a key building block of our
definition of semantic relational validity (i.e.\ the semantic meaning)
of SecRSL judgements (which we present later in \cref{triple}).
It is the relational analogue of RSL's local
validity~\cite[Definition~4]{Vafeiadis_Narayan_13}.
It is defined for a pair of executions $\rchi = \rexec{\res}{\A}{\lab}{\sbo}{\rf}{\mo}{\sco}$ and
$\rchi' = \rexec{\res'}{\A}{\lab'}{\sbo}{\rf}{\mo}{\sco}$ and a
subset~$\V \subseteq \A$ of their actions whose validity is being asserted.
Given heap annotation functions~$\hmap$
and~$\hmap'$ that annotate the $\sbo$ and~$\sw$ edges of
$\rchi$ and~$\rchi'$ respectively with heaps (to witness the memory of each
program at that point during its execution), plus a $\high$ location set~$H$,
local relational validity
asserts that the heap annotations are consistent with the actions $a \in \V$
being
performed at each point in those executions; plus imposing sufficient
conditions to ensure that the executions are isomorphic and
indistinguishable, and that each is safe. It also asserts that all
$\high$ allocations must appear in~$H$ and all $\low$ allocations must not appear in~$H$.

We relegate its definition to \ifExtended \cref{defn:local-valid} in \cref{app}\fi.
For executions~$\rchi = \rexec{\res}{\A}{\lab}{\sbo}{\rf}{\linebreak\mo}{\sco}$ and
$\rchi' = \rexec{\res'}{\A}{\lab'}{\sbo}{\rf}{\mo}{\sco}$,
actions~$\V \subseteq \A$, heap annotation functions~$\hmap$ and~$\hmap'$
and location set~$H$, we write $\Valid{\rchi}{V}{lab'}{\hmap}{\hmap'}{H}$
when local relational validity holds for the subparts of the two executions
identified by~$\V$.

Local relational validity considers just the subparts of the graphs identified
by~$\V$ in order to allow it to be inductively asserted over an ever-increasing
part of the two executions, by the inductive definition of relational validity
of SecRSL judgements. Finally, we now present that definition.

\subsection{Relational Validity}
\label{triple}

We now define the semantic meaning of SecRSL's judgements~$\Triple{HP}{P}{E}{E'}{(y,y').\ Q}{HQ}$. This definition is a novel, compositional and inductive
relational validity property and, to our knowledge, the first such defined
for an axiomatic weak memory semantics. As with the top-level security
properties, we believe its design is of independent interest and should also
be applicable to other axiomatic weak memory models.
However, unlike the top-level definitions, it does not need to be trusted: it is
an internal definition used to carry out the soundness proof of the logic;
not a statement of any guarantees provided by the logic---instead those
guarantees are provided by SecRSL's
adequacy theorem (\cref{thm:adequacy}).

Recall that we refer to the semantic definition that encodes the meaning
of SecRSL's judgements as \emph{relational validity}.  It is defined
formally in \ifExtended \cref{app} as \cref{defn:triple}\fi. Its formal definition uses
various notations introduced in~\citet{Vafeiadis_Narayan_13}; however it is
not necessary to parse the mathematics in order to understand it,
which we carefully explain here.

In order to be compositional, relational validity considers executions of
$E$ and~$E'$ over all possible contexts. That is it considers (what we call)
\emph{contextual executions} in
which the actions of~$E$ are embedded in those of some larger context.
As is common in definitions of
validity for separation logic judgements~\cite{vafeiadismfps11},
relational validity also quantifies
over all possible frame assertions~$R$, to be conjoined with the
precondition~$P$ and postcondition~$Q$, in order to derive the
\textsc{Frame} rule. In addition it quantifies over all possible supersets~$H$
of the initial $\high$-location set~$HP$.

Relational validity of the judgement $\Triple{HP}{P}{E}{E'}{(y,y').\ Q}{HQ}$
considers each
consistent contextual execution of~$E$. For each it considers all potential
contexts of executions of~$E'$ in which the executions
preceding~$E$ and~$E'$
satisfy local relational validity wrt~$H$
and the memories directly preceding~$E$
and~$E'$ satisfy~$P \sep R$. For each it then asserts \emph{configuration
	safety} (\ifExtended \cref{defn:safe} in \cref{app}\fi). Configuration safety
inductively asserts the existence
of an execution of~$E'$ in this context whose final
memories satisfy~$Q \sep R$. It requires that at each step of the
induction the execution of~$E'$ constructed so far satisfies local relational
validity wrt the same part of the execution of~$E$, under the assumption
that the same is true for actions contributed by the contexts.
In this way, configuration safety
guarantees that
the executions obtained when~$E$ and~$E'$ finish are isomorphic\pengbo{I'm not totally sure but I believe this does not hold},
indistinguishable, and that both are safe (i.e.\ free of undefined behaviour
like reads from uninitialised memory or data races, etc.)

In order to define configuration safety, we  had to develop an inductive
characterisation that asserts \emph{partial consistency} of the
execution of~$E'$ constructed at each step of the induction. This property
asserts a subset of the memory model's
consistency axioms and can be applied inductively
to an ever growing subpart of the execution of~$E'$ as it is being constructed.
When combined together with local relational validity, partial consistency
guarantees that for whole programs~$E'$ the entire constructed contextual
execution will be consistent with the axioms of the memory model.
Partial consistency is defined in \ifExtended \cref{app}
(\cref{defn:pcons})\fi.

\subsection{Discussion}\label{sec:discussion}

\paragraph{Soundness Proof}
With the definition of relational validity, the soundness proof of SecRSL
proceeds by proving the soundness of each of the rules against this definition.
This proof follows a similar structure to that of RSL; however is considerably
more complicated because of the additional need to construct the witness
execution of~$E'$.

The resulting Coq proof is ${\sim}21,600$ lines, as compared to
the original Coq proof for RSL which is ${\sim}15,000$ lines.
The two proofs share ${\sim}6,000$ lines in common (basic libraries, C11
language definition and memory model) that, when excluded,
make the SecRSL proof about ${\sim}73\%$ larger.

\paragraph{The $\high$ location set}
Readers familiar with logics for information-flow security might wonder
why SecRSL tracks the locations known to be $\high$ (unobservable to the
attacker). Most other logics instead track which locations are known to
be observable to the attacker, whether through special assertions like
\SecCSL's \emph{location sensitivity assertions}~\cite{Ernst_Murray_19} or
through a static labelling function or otherwise.

The reason is that the sets~$HP$ and~$HQ$ used for this purpose in SecRSL's
judgements~$\Triple{HP}{P}{E}{\linebreak E'}{(y,y).\ Q}{HQ}$ necessarily
\emph{under-approximate} the set of $\high$ locations. They must since
the context in which~$E$ (or~$E'$) executes might have allocated additional
memory locations unused by~$E$.
Were these sets instead used
to track $\low$ locations, they would still need to under-approximate
the true set of attacker-observable locations. However, under-approximating
the set of attacker visible locations is not sound, as it would allow the
logic to ``forget'' that a location was attacker-visible and so allow it to
be written with sensitive data.

\paragraph{Points-To Assertion}
Notice that all points-to assertions generated by SecRSL are of the
form $\Apt{\ell}{\ell}{v}{v'}$ in which the \emph{same} location~$\ell$
is referenced, albeit with two possibly different values~$v$ and~$v'$.
This design choice is intentional and ensures SecRSL's constant-time
guarantee (which requires that \emph{which} memory locations are
accessed by the program and the \emph{order} in which they are accessed
never depends on secrets). We purposefully chose to retain a pair of
locations~$(\ell,\ell)$ to emphasise the relational nature of this assertion.

Indeed,
while SecRSL's rules support compositional
reasoning about partial programs, its
adequacy theorem (\cref{thm:adequacy}) necessarily
applies only to whole programs (a fundamental limitation
it shares with prior separation logics defined for axiomatic memory
models~\cite{Vafeiadis_Narayan_13,FSL,FSLplusplus}
that can give meaningful semantics only to whole programs).
For this reason, modifying SecRSL to attempt to weaken its constant-time
guarantee would not yield a more expressive logic.

\paragraph{Beyond SecRSL}\label{beyond}
We argue explicitly that the ideas underpinning SecRSL's design and
soundness proof
should be readily applicable to other logics that extend the RSL memory model.
Doko and Vafeiadis' Fenced Separation Logic (FSL)~\cite{FSL} is an obvious
target. Indeed an information-flow security analogue of FSL would allow
reasoning about a wider class of programs beyond the Release-Acquire fragment
of C11 considered here.

FSL's memory model is a small extension of
the strengthened memory model of \cref{sec:memory-model} to
add support for fences. A security analogue of FSL would treat fence operations
as potentially attacker observable, much like accesses to atomic locations in
SecRSL (which, recall, must be $\low$). With this insight, SecRSL's adequacy
statement would apply with almost no modification to the FSL semantics.

Just as RSL's atomic location permissions (like $\Aacq{\ell}{Q}$) were
readily adapted to SecRSL's relational setting
while ensuring they behaved identically to their
original counterparts, we conjecture the same should be true for the
additional assertion modalities that FSL introduces for reasoning about
fences.

The definitions of local relational validity (\ifExtended \cref{defn:local-valid}\fi)
could be applied to FSL by simply extending it to add a case for fence
actions, while keeping its current structure. Given the similar structure
of SecRSL's soundness proof to that of RSL, and the close similarity of the
soundness proofs for FSL and RSL, there is strong evidence to suggest that
the soundness proof for a security analogue of FSL should follow a similar
structure to that used in this paper. We leave its development for future
work.

\section{Applying the Logic}\label{applications}

We demonstrate SecRSL by verifying a number of case studies in Coq.

\subsection{Verifying a Spinlock Module}\label{sec:spinlock}

For our first demonstration of SecRSL we show that,
by virtue of its
intentional similarity to RSL, SecRSL allows one to replay RSL proofs
and, in doing so, obtain stronger guarantees than those provable in
RSL.

Specifically we consider the spinlock case study
of~\citet{Vafeiadis_Narayan_13}, which we repeat below and modify only
slightly to ensure that the location~$x$ that is allocated to create the lock
is $\low$.
\begin{align*}
	new\_lock()\defeq&\ \Let{x}{\Alloc{\low}}{\Store{x}{\rel}{1};x}\\
	spin(x)\defeq&\ \Repeat{\Load{x}{\rlx}}\\
	lock(x)\defeq&\ \Repeat{spin(x);\CAS{\acq}{\rlx}{x}{1}{0}} \\
	unlock()\defeq&\ \Store{x}{\rel}{1}
\end{align*}

This module has a SecRSL specification that is almost
identical to its RSL specification.
\begin{align*}
	\{J\}\ &new\_lock()\ \{(x,x). \Lock{x}{J}\}\\
	\{\Lock{x}{J}\}\ &lock(x)\ \{J\sep\Lock{x}{J}\}\\
	\{J\sep\Lock{x}{J}\}\ &unlock(x)\ \{\Lock{x}{J}\}\\
	\Lock{x}{J}\ &\iff \Lock{x}{J} \sep \Lock{x}{J}
\end{align*}

While syntactically identical, note that these specifications once
expressed in SecRSL say that the lock module is not only correct, but also
adheres to SecRSL's constant-time security guarantee.
Thus the lock module will not leak information.

We also observe that these specifications correspond to the rules for
the lock and unlock operations of \SecCSL~\cite{Ernst_Murray_19}. The proof of these specifications
is sketched in \ifExtended \cref{spinlock}, in \cref{app:proofs-spinlock}\fi.

It is no accident that this proof is almost syntactically identical to
its RSL counterpart. This arises not only because the sketch
follows the notational shorthands mentioned on
page~\pageref{shorthands} and in the caption of \cref{fig:mp-proof},
but---more importantly---because we carefully designed SecRSL to
support all of RSL's reasoning principles unchanged. 

\subsection{Mixed-Sensitivity Mutex}\label{sec:spinlock-clas}

A common theme in much recent work on verified information
flow security has been \emph{mixed-sensitivity} (also known as
\emph{value-dependent classification}) data
structures~\cite{Lourenco_Caires_15,Murray_SE_18,Sison_Murray_19,Ernst_Murray_19,frumin2019compositional}. These are ones that can hold data of varying sensitivity over time,
where the data structure maintains information about the sensitivity
of the data it currently contains.

We demonstrate SecRSL's ability to reason about the implementations of
such concurrent data structures for the first time, while taking account of
weak memory effects. We extend the prior spinlock module so that it protects
access to a mixed-sensitivity memory location. We refer to this resulting
abstraction as a \emph{mixed-sensitivity mutex}.

\newcommand{\lockMSM}[1]{\mathsf{lock\_MSM}(#1)}
\newcommand{\unlockMSML}[1]{\mathsf{unlock\_MSM\_L}(#1)}
\newcommand{\unlockMSMH}[1]{\mathsf{unlock\_MSM\_H}(#1)}
\newcommand{\newMSM}{\mathsf{new\_MSM}()}
$\newMSM$ creates a mixed sensitivity mutex~$a$.
The $\lockMSM{a}$ operation acquires access to the mixed-sensitivity location and
returns a boolean indicating whether it currently holds sensitive ($\high$)
data or not. The module provides two operations, $\unlockMSMH{a}$ and $\unlockMSML{a}$,
to relinquish access to the
mixed-sensitivity location, depending on whether the data it now contains is
$\high$ or $\low$ respectively.

\begin{figure}
  \begin{tabular}{ll}
    \begin{minipage}{0.4\textwidth}
	$\newMSM\ \ \defeq$ \\
	\ \ \ \ \begin{tabular}{l}
		$\Let{b}{\Alloc{\high}}{}$\\
		$\Let{a}{\Alloc{\low}}{}$\\
		$\Store{b}{\na}{0};$\\
	$\Store{a}{\rel}{1};$\\
        \end{tabular}
    \end{minipage} &
        
    \begin{minipage}{0.4\textwidth}
	$\lockMSM{a}\ \ \defeq$ \\
	\ \ \ \ \begin{tabular}{l}
		$\mathbf{repeat}$\\
		$\quad\Let{x}{(\Repeat{\Load{a}{\rlx}})}{}$\\
		$\quad\Let{y}{\CAS{\acq}{\rlx}{a}{x}{0}}{}$\\
		$\quad \If{(x==y)}{x}{0}$\\
	$\mathbf{end}$\\
        \end{tabular}
     \end{minipage}
  \end{tabular}\\

	\hspace*{-1.5cm}$\unlockMSML{a}\ \ \defeq\ \ \Store{a}{\rel}{1}$\qquad\qquad\quad
	$\unlockMSMH{a}\ \ \defeq\ \ \Store{a}{\rel}{2}$
        \caption{The mixed-sensitivity mutex implementation.\label{fig:spinlock-clas}}
\end{figure}

\cref{fig:spinlock-clas} shows its implementation. The original
spinlock writes 0 to location~$a$ to indicate that the lock is
occupied (acquired) and 1 for unoccupied (free). The mixed-sensitivity mutex
uses 0 to indicate that the mutex is occupied, while 1 and 2 indicate it is
unoccupied and holding $\low$ (1) or $\high$ (2) data.

Defining the predicates $\CLow{\ell}$ (respectively $\CHigh{\ell}$) to denote when
(non-atomic) location
$\ell$ holds a possibly $\high$ (respectively definitely $\low$) value,
we verify the mixed-sensitivity mutex against the following specifications.
Here the predicate $\LockC{a}$ says that location~$a$ refers to a
mixed-sensitivity mutex.

\begin{center}
	$\Triple{}{\Aemp}{\newMSM}{\newMSM}{\LockC{a}}{b}$\\[0.5ex]
	$\LockC{a}\iff\LockC{a}\sep\LockC{a}$\\[0.5ex]
	$\TripleAH{b}{\LockC{a}\sep\CLow{b}}{\unlockMSML{a}}{\LockC{a}}{b}$\\[0.5ex]
	$\TripleAH{b}{\LockC{a}\sep\CHigh{b}}{\unlockMSMH{a}}{\LockC{a}}{b}$\\[0.5ex]
	$\TripleAH{b}{\LockC{a}}{\lockMSM{a}}{(y,y).\ \LockC{a}\sep(\metaITE{y=1}{\CLow{b}}{\CHigh{b}})}{b}$\\
\end{center}

The first says that $\newMSM$ creates mixed-sensitivity mutexes;
the second that $\LockC{a}$ is freely duplicable; the third requires that
when unlocking with the $\unlockMSML{a}$ operation, that $a$ holds a $\low$
value ($\CLow{a}$); the final specification says that after locking,
the (necessarily $\low$) return value~$y$ correctly indicates the sensitivity
of the data held in location~$a$.

The proof sketch for this example appears in \ifExtended \cref{app:proofs-spinlock-clas}\fi.

\subsection{Implementing Verified Synchronous Channels}\label{sec:syncchannel}
Message-passing concurrency, in which concurrent threads transfer
data over \emph{channels} rather than (raw) shared memory, is a common
programming abstraction and has been widely studied in the context of 
information-flow security~\cite{allen1991comparison,roscoe1994non,roscoe1995csp,honda2000secure,zdancewic2003observational,terauchi2008type,Murray_Lowe_10,Karbyshev+:POST18}.

One of the most common message-passing abstractions is the
\emph{synchronous channel}, e.g.\ as widely studied in various process
calculi like CSP, CCS, and the synchronous $\pi$-calculus,
and implemented in various programming languages like Go and generalised
by Ada's rendezvous mechanism.

A synchronous channel allows data to be transmitted from a sending thread
to a receiving thread. Both threads block until the other is ready, which
means that this type of channel also forces the sender and receiver to
synchronise each time that data is transmitted on the channel.

\newcommand{\newchannel}{\mathsf{new\_channel()}}
\newcommand{\send}[2]{\mathsf{send}(#1,#2)}
\newcommand{\recv}[1]{\mathsf{recv}(#1)}
\newcommand{\ishigh}{\mathit{ishigh}}
\newcommand{\isH}{\mathit{isH}}

\subsubsection{Release/Acquire Synchronous Channel}\label{sec:syncchannel-ra}

We implemented in C, and verified in SecRSL in Coq,  a synchronous channel
abstraction which is specifically designed to support transmitting
data of varying sensitivity. The (inline) $\newchannel$ operation creates
a new channel; $\send{v}{\ishigh}$ synchronously
sends the value~$v$ on the channel, where~$v$'s sensitivity is given by the boolean~$\ishigh$; $\recv{d}$ takes a
pointer argument~$d$ and waits to receive the next value from the channel,
which is written to location~$d$ and whose sensitivity is returned as
the (boolean) return-value of $\recv{\,}$.

The code for the synchronous channel appears in \cref{fig:syncchannel}.
The verified $\newchannel$ operation yields distinct permissions (in the
form of SecRSL predicates) to send and receive on the channel.
The $\newchannel$ operation creates three locations: an atomic location~$a$
and two non-atomic locations~$b$ and~$c$. Location~$c$ is used only by the
receiver. The sender begins with permission to
perform a release write to location~$a$, and ownership of non-atomic
location~$b$. The receiver has permission to
perform an acquire read to~$a$ and owns the non-atomic location~$c$.

A synchronous communication on the channel 
involves the sender writing to~$b$ the value~$v$ to be transmitted and
then performing a release write to~$a$ to encode the sensitivity of~$v$
(similarly to the message-passing program in \cref{fig:mp}). In doing so,
the sender transfers ownership of~$b$ to the receiver. The sender then
busywaits for the receiver to
return the ownership to the sender (so that it can be used to perform
subsequent sends).

The receiver does so after waiting to receive on the channel, which involves
polling~$a$ via acquire reads until its value changes. At this point the
receiver has acquired ownership of~$b$ to perform a release
write to~$a$. The receiver reads~$b$ to learn the value~$v$ that was
transmitted, and infers $v$'s sensitivity from the value it read from~$a$.
Finally it releases the ownership it got from the sender by performing
a release write to~$a$.

Thus each send/receive pair involves a two-way transfer of
location $b$,
neatly illustrating SecRSL's power for reasoning about
concurrency abstractions involving ownership transfer.

To learn when $a$'s value changes, the receiver keeps a local copy in
location~$c$ of
the most recent value that the receiver wrote to~$a$.
The sender encodes $v$'s sensitivity ($\ishigh$) by incrementing~$a$ either
by 2 (if $\ishigh$ is true) or 1 (otherwise). The receiver releases
the location it got from the sender by incrementing~$a$ by 3 from its
original value (before it was modified by the sender). Thus the sender owns
$b$ whenever $a$'s value is divisible by 3; the receiver owns~$b$ otherwise,
in which case its sensitivity is determined by the (nonzero)
value of $\Mod{n}{3}$.

\begin{figure}
\begin{minipage}{0.9\textwidth}
  \begin{tabular}{ll}
      \begin{minipage}{0.4\textwidth}
	$\newchannel\ \ \defeq\ \ $ \\
	\ \ \ \ \begin{tabular}{l}
			$\Let{b}{\Alloc{\high}}{}$\\
			$\Let{c}{\Alloc{\low}}{}$\\
			$\Let{a}{\Alloc{\low}}{}$\\
			$\Store{b}{\na}{0};$\\
			$\Store{c}{\na}{0};$\\
			$\Store{a}{\rel}{0};$\\
	\end{tabular}
      \end{minipage} &
      \begin{minipage}{0.4\textwidth}
	$\send{v}{\ishigh}\ \ \defeq$ \\
	\ \ \ \ \begin{tabular}{l}
		$\Let{x}{\Load{b}{\na}}{}$\\
		$\Store{b}{\na}{v};$\\
		$\Store{a}{\rel}{x+(\ishigh\ ?\ 2:1)};$\\
		$\mathbf{repeat}$\\
		$\quad \Let{z}{\Load{a}{\acq}}{}$\\
		$\quad \quad \If{z==x+3}{1}{0}$\\
		$\mathbf{end}$\\
		$\Store{b}{\na}{x+3};$\\
	\end{tabular}
      \end{minipage}
    \end{tabular}
    \ \\
	$\recv{d}\ \ \defeq$ \\
	\begin{tabular}{l}
		$\Let{t}{\Load{c}{\na}}{}$\\
		$\mathbf{let}\ lv\ =\ \mathbf{(repeat}$\\
		$\qquad \qquad \qquad \Let{z}{\Load{a}{\acq}}{}$\\
		$\qquad \qquad \qquad \quad \If{(z==t+1)||(z==t+2)}{z}{0}$\\
		$\qquad \qquad \quad \ \mathbf{end)}\ \mathbf{in}$\\
		$\Let{v}{\Load{b}{\na}}{}$\\
		$\Store{d}{\na}{v};$\\
		$\Store{a}{\rel}{t+3};$\\
		$\Store{c}{\na}{t+3};$\\
		$lv-t;$\\
	\end{tabular}
\end{minipage}
\caption{The synchronous channel implementation.\label{fig:syncchannel} Here
``$||$'' denotes boolean ``or'' (disjunction).}
\end{figure}

The permission to send on the channel is encoded in the $\Sender{n}$ predicate, where~$n$
denotes the value currently stored in location~$b$, the sender's counter. Likewise, $\Recver{n}$
is the permission to receive, where $n$ records the current value stored in~$c$, the receiver's counter.
Reusing the predicates $\CLow{\ell}$ and $\CHigh{\ell}$ from
\cref{sec:spinlock-clas} to denote when
(non-atomic) location
$\ell$ holds a possibly-$\high$ (respectively definitely $\low$) value, we prove
the following SecRSL specifications,
for all $n$ such that $\Mod{n}{3}$ = 0:

\begin{center}
	$\Triple{}{\Aemp}{\linebreak\newchannel}{\newchannel\linebreak}{\ \Sender{0}\ \sep\ \Recver{0}\ }{b}$\\
	\ \\
	$\Triple{b}{\ \Sender{n}\ \sep (\metaITE{\ishigh}{\Aemp}{\Alow{v}{v'}})\ }{\linebreak\send{v}{\ishigh}}{\send{v'}{\ishigh}\linebreak}{\ \Sender{n+3}\ }{b}$\\
	\ \\
	$\Triple{b,d}{\ \Recver{n}\ \sep (\CHigh{d})\ }{\linebreak\recv{d}}{\recv{d}\linebreak}{(y,y').\ \Recver{n+3}\ \sep (\metaITE{y=1}{\CLow{d}}{\CHigh{d}}) \sep \ldots\linebreak \land (y=y')\land(y=1 \lor y=2)}{b,d}$\\
\ \\
$\CLow{\ell}\  \defeq\ \Aexists{v}{v'}{\Apt{\ell}{\ell}{v}{v'} \sep \Alow{v}{v'}}$\\
$\CHigh{\ell}\  \defeq\ \Aexists{v}{v'}{\Apt{\ell}{\ell}{v}{v'}}$
\end{center}

These say that $\newchannel$ yields the permissions to send and receive,
initialising the locations to 0. To call $\send{v}{\ishigh}$, we must have the
sender permission and $v$'s classification must match~$\ishigh$.  After
sending, the permission to subsequently send is returned to allow the sender
to send repeatedly. The specification for $\recv{\,}$ has a
similar structure and additionally requires the location~$d$, where the
value received will be written, is valid. The ``\ldots'' in the
postcondition for $\recv{\,}$ elides additional permissions that
the logic tracks: namely the now-useless permission for $\recv$ to
read the alternative value ($n+1$ or $n+2$) from~$a$, other than the
value it \emph{did} read and that will now never be written to~$a$
(see \ifExtended \cref{app:syncchannel-proof}\fi).
Of course one can always safely ignore the additional
permission in proofs by employing the frame rule.

The proof sketch appears in \ifExtended \cref{app:syncchannel-proof}\fi.

\subsubsection{Release/CAS Synchronous Channel}\label{sec:syncchannel-cas}
To showcase SecRSL's flexibility, we also implemented and
verified an alternative synchronous channel implementation. Rather than
using release/acquire synchronisation, this implementation instead
uses release/CAS pairs. Doing so avoids the need for the counter~$n$
and leads to a simpler proof, albeit with a similar structure to the
original.  This implementation we also verified in Coq, against the
following specifications, where $\SenderC$ and $\RecverC$ denote the
permissions to send and receive respectively:
\newcommand{\newchannelCAS}{\mathsf{new\_channel\_CAS}()}
\newcommand{\sendCAS}[2]{\mathsf{send\_CAS}(#1,#2)}
\newcommand{\recvCAS}[1]{\mathsf{recv\_CAS}(#1)}
\begin{center}
	$\Triple{}{\Aemp}{\linebreak\newchannelCAS}{\newchannelCAS\linebreak}{\ \SenderC\ \sep\ \RecverC\ }{b}$\\
	\ \\
	$\Triple{b}{\ \SenderC\ \sep (\metaITE{\ishigh}{\Aemp}{\Alow{v}{v'}})\ }{\linebreak\sendCAS{v}{\ishigh}}{\sendCAS{v'}{\ishigh}\linebreak}{\ \SenderC\ }{b}$\\
	\ \\
	$\Triple{b,d}{\ \RecverC\ \sep (\CHigh{d})\ }{\linebreak\recvCAS{d}}{\recvCAS{d}\linebreak}{(y,y').\ \RecverC\ \sep (\metaITE{y=1}{\CLow{d}}{\CHigh{d}}) \linebreak \land (y=y') \land (y=1 \lor y=2)}{b,d}$\\
\end{center}

\subsubsection{Performance Comparison}\label{sec:eval}
We implemented both verified channel implementations in C, and performed
a rudimentary comparison of their performance. As a baseline we implemented
a sequentially-consistent version of the synchronous channel, shown in
\ifExtended \cref{fig:baseline} (in \cref{app:baseline})\fi. We constructed this implementation by taking the
one from \cref{fig:syncchannel} and marking all shared variable accesses 
with the $\scx$ mode, in order to make its execution sequentially consistent
(i.e.\ insulate it from weak memory effects). While this
implementation is not verified, we posit that it might be verifiable using
a suitable adaptation of the fine-grained security separation logic of~\citet{frumin2019compositional}.

The results are summarised in \cref{tbl:syncchannel-performance}. We report
maximum and minimum average bandwidths observed when transferring 1 GiB
(1024 MiB) of random data over four trials, across various platforms. We also report the
average number of rounds per second observed across all trials for each platform. The results include a version of the release/acquire channel implementation
that transmits 8-byte values (\textsf{long int}s); all other implementations
transmit 4-byte values (\textsf{int}s).







\begin{table}
  \begin{tabular}{ c|c|c|c }
                               & \multicolumn{3}{c}{Observed Bandwidth MiB/sec (avg.\ rounds/sec)} \\
		Implementation & Intel MacBook 
		         & Intel Server  
		& ARM Server  \\
                \hline\hline
		Rel/Acq (4 bytes) & 31--33 ($8.3 \times 10^6$) & 17--19 ($4.7 \times 10^6$) & 13--19 ($4.1 \times 10^6$) \\ 
		Rel/Acq (8 bytes) & 60--65 ($8.3 \times 10^6$) & 33--36 ($4.6 \times 10^6$) & 23--26 ($3.2 \times 10^6$) \\ 
		Rel/CAS (4 bytes) & 15--18 ($4.2 \times 10^6$) & 13--15 ($3.7 \times 10^6$) & 9--12 ($2.7 \times 10^6$) \\ 
		SC (4 bytes) & 17--18 ($4.8 \times 10^6$) & 9.2--9.6 ($2.5 \times 10^6$) & 14--19 ($4.2 \times 10^6$) \\ 
		\hline
	\end{tabular}
        \caption{Performance comparison of the verified synchronous channel implementations.\label{tbl:syncchannel-performance}  Platforms: \emph{Intel MacBook}: MacBook 13-inch, 2020, 2 GHz Quad-Core Intel Core i5, MacOS. \emph{Intel Server}:  8 CPU Xeon(R) Gold 6248 @ 2.50GHz, Linux x86\_64 on VMWare hypervisor with full visualisation. \emph{ARM Server}:  2 CPU Neoverse-N1, Linux ARM64 (aarch64), Amazon AWS EC2 t4g.micro instance.}
\end{table}

The release/acquire implementation consistently outperforms the others on
both Intel platforms, with observed throughput improvements of up to 88\%.
Naturally doubling the size of the quantity transferred in
each round tends to double the observed bandwidth.
On ARMv8 AArch64, $\acq$ loads and $\scx$ loads both map to the \texttt{LDAR}
Load-Acquire instruction; likewise $\rel$ and $\scx$ stores both map to the
\texttt{STLR} Store-Release instruction~\cite{Sewell:mappings}.
We conjecture that this is why the release/acquire
version was observed to perform no better on the ARM platform than
the sequentially-consistent baseline. Indeed, the latter
was observed to perform slightly better
than the former; however, the results on this platform show
considerable variability and this observed difference is well within the
noise. 

These empirical results demonstrate that, depending on the deployment platform,
significant performance improvements
can be obtained by utilising the C11 weak memory primitives, backed by the
formal guarantees afforded by SecRSL. They clearly demonstrate the power of
SecRSL over prior logics like SecCSL~\cite{Ernst_Murray_19}  which can reason
only about data-race free programs, or even more recent logics~\cite{frumin2019compositional} that implicitly assume sequential consistency. 

\section{Further Related Work}

Prior security logics and type systems for information-flow security
on weak memory models
include that of~\citet{vaughan2012secure} who
developed a simple security type system for an operational semantics
of the TSO memory model, and
\citet{mantel2014noninterference} who developed
a transforming type system for ensuring security also on PSO and the IBM370.
As security type systems, neither supported the
precision afforded by a logic like SecRSL.

\citet{smith2019value} present a program logic for proving
secure information flow of ARMv8. Unlike SecRSL which targets C11,
they target a low level memory model.  Like SecRSL, their logic supports
reasoning about value-dependent classification. However, unlike SecRSL,
theirs does not support local reasoning with ownership transfer and invariants.
As \cref{fig:mp-proof\ifExtended,spinlock\fi} demonstrates, such support is vital for practical
reasoning about expressive security policies.

Our security definitions over the C11 axiomatic memory model are an instance
of noninterference~\cite{Goguen_Meseguer_82} for a
so-called ``true concurrency'' semantics. It would be interesting to
compare how our definitions relate to those for other true concurrency
models, e.g.~those
for Petri Nets~\cite{baldan2014non,baldan2018multilevel}.\\

\section{Conclusion}
We presented SecRSL, a security separation logic for C11 Release-Acquire
concurrency. SecRSL inherits RSL's
virtues of compositional, local reasoning about Release-Acquire
atomics, plus \SecCSL's ability to reason about expressive security
  policies like value-dependent classification.

  We also presented
  the first definition of information-flow security for an
  axiomatic weak memory model, against which we proved SecRSL sound.
  SecRSL ensures that programs satisfy a constant-time security guarantee,
  while being free of undefined behaviour.

  We demonstrated SecRSL by using it to implement and verify
the functional correctness and 
security of various concurrency primitives, including a spinlock
module, a mixed-sensitivity mutex, and two synchronous channel
implementations. Benchmarking the latter against an unverified
sequentially-consistent
implementation showed that SecRSL can enable significant
performance gains.

Beyond these examples, we also believe
  (as \cref{beyond} argues)
  these ideas are of interest---and can be readily applied---beyond the
  Release-Acquire fragment of C11.

\begin{DIFnomarkup}
\begin{acks}                            
  We thank the anonymous reviewers for their insightful feedback on earlier
  drafts of this paper. 
  This material is based upon work supported by the
  Commonwealth of Australia Defence Science and
  Technology Group, Next Generation Technologies Fund (NGTF).
\end{acks}
\end{DIFnomarkup}

\bibliographystyle{ACM-Reference-Format}
\bibliography{references}


\begin{thebibliography}{42}


\ifx \showCODEN    \undefined \def \showCODEN     #1{\unskip}     \fi
\ifx \showDOI      \undefined \def \showDOI       #1{#1}\fi
\ifx \showISBNx    \undefined \def \showISBNx     #1{\unskip}     \fi
\ifx \showISBNxiii \undefined \def \showISBNxiii  #1{\unskip}     \fi
\ifx \showISSN     \undefined \def \showISSN      #1{\unskip}     \fi
\ifx \showLCCN     \undefined \def \showLCCN      #1{\unskip}     \fi
\ifx \shownote     \undefined \def \shownote      #1{#1}          \fi
\ifx \showarticletitle \undefined \def \showarticletitle #1{#1}   \fi
\ifx \showURL      \undefined \def \showURL       {\relax}        \fi
\providecommand\bibfield[2]{#2}
\providecommand\bibinfo[2]{#2}
\providecommand\natexlab[1]{#1}
\providecommand\showeprint[2][]{arXiv:#2}

\bibitem[\protect\citeauthoryear{Alglave, Fox, Ishtiaq, Myreen, Sarkar, Sewell,
  and Nardelli}{Alglave et~al\mbox{.}}{2009}]%
        {alglave2009semantics}
\bibfield{author}{\bibinfo{person}{Jade Alglave}, \bibinfo{person}{Anthony
  Fox}, \bibinfo{person}{Samin Ishtiaq}, \bibinfo{person}{Magnus~O Myreen},
  \bibinfo{person}{Susmit Sarkar}, \bibinfo{person}{Peter Sewell}, {and}
  \bibinfo{person}{Francesco~Zappa Nardelli}.} \bibinfo{year}{2009}\natexlab{}.
\newblock \showarticletitle{The semantics of {Power} and {ARM} multiprocessor
  machine code}. In \bibinfo{booktitle}{\emph{Proceedings of the 4th workshop
  on Declarative aspects of multicore programming}}. \bibinfo{pages}{13--24}.
\newblock


\bibitem[\protect\citeauthoryear{Allen}{Allen}{1991}]%
        {allen1991comparison}
\bibfield{author}{\bibinfo{person}{PG Allen}.} \bibinfo{year}{1991}\natexlab{}.
\newblock \showarticletitle{A comparison of non-interference and
  non-deducibility using {CSP}}. In \bibinfo{booktitle}{\emph{IEEE Computer
  Security Foundations Workshop (CSFW)}}. IEEE, \bibinfo{pages}{43--54}.
\newblock


\bibitem[\protect\citeauthoryear{Baldan and Beggiato}{Baldan and
  Beggiato}{2018}]%
        {baldan2018multilevel}
\bibfield{author}{\bibinfo{person}{Paolo Baldan} {and}
  \bibinfo{person}{Alessandro Beggiato}.} \bibinfo{year}{2018}\natexlab{}.
\newblock \showarticletitle{Multilevel transitive and intransitive
  non-interference, causally}.
\newblock \bibinfo{journal}{\emph{Theoretical Computer Science}}
  \bibinfo{volume}{706} (\bibinfo{year}{2018}), \bibinfo{pages}{54--82}.
\newblock


\bibitem[\protect\citeauthoryear{Baldan and Carraro}{Baldan and
  Carraro}{2014}]%
        {baldan2014non}
\bibfield{author}{\bibinfo{person}{Paolo Baldan} {and} \bibinfo{person}{Alberto
  Carraro}.} \bibinfo{year}{2014}\natexlab{}.
\newblock \showarticletitle{Non-interference by unfolding}. In
  \bibinfo{booktitle}{\emph{International Conference on Applications and Theory
  of Petri Nets and Concurrency}}. Springer, \bibinfo{pages}{190--209}.
\newblock


\bibitem[\protect\citeauthoryear{Barthe, Blazy, Gr{\'e}goire, Hutin, Laporte,
  Pichardie, and Trieu}{Barthe et~al\mbox{.}}{2019}]%
        {barthe2019formal}
\bibfield{author}{\bibinfo{person}{Gilles Barthe}, \bibinfo{person}{Sandrine
  Blazy}, \bibinfo{person}{Benjamin Gr{\'e}goire}, \bibinfo{person}{R{\'e}mi
  Hutin}, \bibinfo{person}{Vincent Laporte}, \bibinfo{person}{David Pichardie},
  {and} \bibinfo{person}{Alix Trieu}.} \bibinfo{year}{2019}\natexlab{}.
\newblock \showarticletitle{Formal verification of a constant-time preserving C
  compiler}.
\newblock \bibinfo{journal}{\emph{Proceedings of the ACM on Programming
  Languages}} \bibinfo{volume}{4}, \bibinfo{number}{POPL}
  (\bibinfo{year}{2019}), \bibinfo{pages}{1--30}.
\newblock


\bibitem[\protect\citeauthoryear{Barthe, Espitau, Gr{\'e}goire, Hsu, and
  Strub}{Barthe et~al\mbox{.}}{2017}]%
        {barthe2017proving}
\bibfield{author}{\bibinfo{person}{Gilles Barthe}, \bibinfo{person}{Thomas
  Espitau}, \bibinfo{person}{Benjamin Gr{\'e}goire}, \bibinfo{person}{Justin
  Hsu}, {and} \bibinfo{person}{Pierre-Yves Strub}.}
  \bibinfo{year}{2017}\natexlab{}.
\newblock \showarticletitle{Proving expected sensitivity of probabilistic
  programs}.
\newblock \bibinfo{journal}{\emph{Proceedings of the ACM on Programming
  Languages}} \bibinfo{volume}{2}, \bibinfo{number}{POPL}
  (\bibinfo{year}{2017}), \bibinfo{pages}{1--29}.
\newblock


\bibitem[\protect\citeauthoryear{Batty, Donaldson, and Wickerson}{Batty
  et~al\mbox{.}}{2016}]%
        {batty2016overhauling}
\bibfield{author}{\bibinfo{person}{Mark Batty}, \bibinfo{person}{Alastair~F
  Donaldson}, {and} \bibinfo{person}{John Wickerson}.}
  \bibinfo{year}{2016}\natexlab{}.
\newblock \showarticletitle{Overhauling {SC} atomics in {C11} and {OpenCL}}. In
  \bibinfo{booktitle}{\emph{ACM SIGPLAN-SIGACT Symposium on Principles of
  Programming Languages (POPL)}}. \bibinfo{pages}{634--648}.
\newblock


\bibitem[\protect\citeauthoryear{Batty, Owens, Sarkar, Sewell, and Weber}{Batty
  et~al\mbox{.}}{2011}]%
        {batty2011mathematizing}
\bibfield{author}{\bibinfo{person}{Mark Batty}, \bibinfo{person}{Scott Owens},
  \bibinfo{person}{Susmit Sarkar}, \bibinfo{person}{Peter Sewell}, {and}
  \bibinfo{person}{Tjark Weber}.} \bibinfo{year}{2011}\natexlab{}.
\newblock \showarticletitle{Mathematizing {C++} concurrency}. In
  \bibinfo{booktitle}{\emph{ACM SIGPLAN-SIGACT Symposium on Principles of
  Programming Languages (POPL)}}. \bibinfo{pages}{55--66}.
\newblock


\bibitem[\protect\citeauthoryear{Batty}{Batty}{2014}]%
        {Batty:phd}
\bibfield{author}{\bibinfo{person}{Mark~John Batty}.}
  \bibinfo{year}{2014}\natexlab{}.
\newblock \emph{\bibinfo{title}{The {C11} and {C++11} Concurrency Model}}.
\newblock \bibinfo{thesistype}{Ph.D. Dissertation}. \bibinfo{school}{University
  of Cambridge}.
\newblock


\bibitem[\protect\citeauthoryear{Benton}{Benton}{2004}]%
        {Benton_04}
\bibfield{author}{\bibinfo{person}{Nick Benton}.}
  \bibinfo{year}{2004}\natexlab{}.
\newblock \showarticletitle{Simple relational correctness proofs for static
  analyses and program transformations}. In \bibinfo{booktitle}{\emph{ACM
  SIGPLAN-SIGACT Symposium on Principles of Programming Languages (POPL)}}.
  \bibinfo{pages}{14--25}.
\newblock


\bibitem[\protect\citeauthoryear{Doko and Vafeiadis}{Doko and
  Vafeiadis}{2016}]%
        {FSL}
\bibfield{author}{\bibinfo{person}{Marko Doko} {and} \bibinfo{person}{Viktor
  Vafeiadis}.} \bibinfo{year}{2016}\natexlab{}.
\newblock \showarticletitle{A program logic for {C11} memory fences}. In
  \bibinfo{booktitle}{\emph{International Conference on Verification, Model
  Checking, and Abstract Interpretation (VMCAI)}}. Springer,
  \bibinfo{pages}{413--430}.
\newblock


\bibitem[\protect\citeauthoryear{Doko and Vafeiadis}{Doko and
  Vafeiadis}{2017}]%
        {FSLplusplus}
\bibfield{author}{\bibinfo{person}{Marko Doko} {and} \bibinfo{person}{Viktor
  Vafeiadis}.} \bibinfo{year}{2017}\natexlab{}.
\newblock \showarticletitle{Tackling real-life relaxed concurrency with
  {FSL}++}. In \bibinfo{booktitle}{\emph{European Symposium on Programming
  (ESOP)}}. Springer, \bibinfo{pages}{448--475}.
\newblock


\bibitem[\protect\citeauthoryear{Ernst and Murray}{Ernst and Murray}{2019}]%
        {Ernst_Murray_19}
\bibfield{author}{\bibinfo{person}{Gidon Ernst} {and} \bibinfo{person}{Toby
  Murray}.} \bibinfo{year}{2019}\natexlab{}.
\newblock \showarticletitle{\textsc{SecCSL:} Security Concurrent Separation
  Logic}. In \bibinfo{booktitle}{\emph{International Conference on Computer
  Aided Verification (CAV)}}. \bibinfo{pages}{208--230}.
\newblock


\bibitem[\protect\citeauthoryear{Flanagan, Sabry, Duba, and Felleisen}{Flanagan
  et~al\mbox{.}}{1993}]%
        {flanagan1993essence}
\bibfield{author}{\bibinfo{person}{Cormac Flanagan}, \bibinfo{person}{Amr
  Sabry}, \bibinfo{person}{Bruce~F Duba}, {and} \bibinfo{person}{Matthias
  Felleisen}.} \bibinfo{year}{1993}\natexlab{}.
\newblock \showarticletitle{The essence of compiling with continuations}. In
  \bibinfo{booktitle}{\emph{ACM SIGPLAN Conference on Programming Language
  Design and Implementation (PLDI)}}. \bibinfo{pages}{237--247}.
\newblock


\bibitem[\protect\citeauthoryear{Frumin, Krebbers, and Birkedal}{Frumin
  et~al\mbox{.}}{2021}]%
        {frumin2019compositional}
\bibfield{author}{\bibinfo{person}{Dan Frumin}, \bibinfo{person}{Robbert
  Krebbers}, {and} \bibinfo{person}{Lars Birkedal}.}
  \bibinfo{year}{2021}\natexlab{}.
\newblock \showarticletitle{Compositional Non-Interference for Fine-Grained
  Concurrent Programs}. In \bibinfo{booktitle}{\emph{IEEE Symposium on Security
  \& Privacy (S\&P)}}.
\newblock
\newblock
\shownote{To appear.}


\bibitem[\protect\citeauthoryear{Goguen and Meseguer}{Goguen and
  Meseguer}{1982}]%
        {Goguen_Meseguer_82}
\bibfield{author}{\bibinfo{person}{Joseph Goguen} {and}
  \bibinfo{person}{Jos{\'e} Meseguer}.} \bibinfo{year}{1982}\natexlab{}.
\newblock \showarticletitle{Security Policies and Security Models}. In
  \bibinfo{booktitle}{\emph{IEEE Symposium on Security \& Privacy (S\&P)}}.
  \bibinfo{publisher}{IEEE Computer Society}, \bibinfo{address}{Oakland,
  California, USA}, \bibinfo{pages}{11--20}.
\newblock


\bibitem[\protect\citeauthoryear{Honda, Vasconcelos, and Yoshida}{Honda
  et~al\mbox{.}}{2000}]%
        {honda2000secure}
\bibfield{author}{\bibinfo{person}{Kohei Honda}, \bibinfo{person}{Vasco
  Vasconcelos}, {and} \bibinfo{person}{Nobuko Yoshida}.}
  \bibinfo{year}{2000}\natexlab{}.
\newblock \showarticletitle{Secure information flow as typed process
  behaviour}. In \bibinfo{booktitle}{\emph{European Symposium on Programming
  (ESOP)}}. Springer, \bibinfo{pages}{180--199}.
\newblock


\bibitem[\protect\citeauthoryear{Karbyshev, Svendsen, Askarov, and
  Birkedal}{Karbyshev et~al\mbox{.}}{2018}]%
        {Karbyshev+:POST18}
\bibfield{author}{\bibinfo{person}{Aleksandr Karbyshev},
  \bibinfo{person}{Kasper Svendsen}, \bibinfo{person}{Aslan Askarov}, {and}
  \bibinfo{person}{Lars Birkedal}.} \bibinfo{year}{2018}\natexlab{}.
\newblock \showarticletitle{Compositional Non-Interference for Concurrent
  Programs via Separation and Framing}. In
  \bibinfo{booktitle}{\emph{International Conference on Principles of Security
  and Trust (POST)}}.
\newblock


\bibitem[\protect\citeauthoryear{Louren\c{c}o and Caires}{Louren\c{c}o and
  Caires}{2015}]%
        {Lourenco_Caires_15}
\bibfield{author}{\bibinfo{person}{Lu\'{i}sa Louren\c{c}o} {and}
  \bibinfo{person}{Lu\'{i}s Caires}.} \bibinfo{year}{2015}\natexlab{}.
\newblock \showarticletitle{Dependent Information Flow Types}. In
  \bibinfo{booktitle}{\emph{ACM SIGPLAN-SIGACT Symposium on Principles of
  Programming Languages (POPL)}}. \bibinfo{address}{Mumbai, India},
  \bibinfo{pages}{317--328}.
\newblock


\bibitem[\protect\citeauthoryear{Mador-Haim, Maranget, Sarkar, Memarian,
  Alglave, Owens, Alur, Martin, Sewell, and Williams}{Mador-Haim
  et~al\mbox{.}}{2012}]%
        {mador2012axiomatic}
\bibfield{author}{\bibinfo{person}{Sela Mador-Haim}, \bibinfo{person}{Luc
  Maranget}, \bibinfo{person}{Susmit Sarkar}, \bibinfo{person}{Kayvan
  Memarian}, \bibinfo{person}{Jade Alglave}, \bibinfo{person}{Scott Owens},
  \bibinfo{person}{Rajeev Alur}, \bibinfo{person}{Milo~MK Martin},
  \bibinfo{person}{Peter Sewell}, {and} \bibinfo{person}{Derek Williams}.}
  \bibinfo{year}{2012}\natexlab{}.
\newblock \showarticletitle{An axiomatic memory model for {POWER}
  multiprocessors}. In \bibinfo{booktitle}{\emph{International Conference on
  Computer Aided Verification (CAV)}}. \bibinfo{pages}{495--512}.
\newblock


\bibitem[\protect\citeauthoryear{Maillard, Hri{\c{t}}cu, Rivas, and
  Van~Muylder}{Maillard et~al\mbox{.}}{2019}]%
        {maillard2019next}
\bibfield{author}{\bibinfo{person}{Kenji Maillard},
  \bibinfo{person}{C{\u{a}}t{\u{a}}lin Hri{\c{t}}cu}, \bibinfo{person}{Exequiel
  Rivas}, {and} \bibinfo{person}{Antoine Van~Muylder}.}
  \bibinfo{year}{2019}\natexlab{}.
\newblock \showarticletitle{The next 700 relational program logics}.
\newblock \bibinfo{journal}{\emph{Proceedings of the ACM on Programming
  Languages}} \bibinfo{volume}{4}, \bibinfo{number}{POPL}
  (\bibinfo{year}{2019}), \bibinfo{pages}{1--33}.
\newblock


\bibitem[\protect\citeauthoryear{Manson, Pugh, and Adve}{Manson
  et~al\mbox{.}}{2005}]%
        {manson2005java}
\bibfield{author}{\bibinfo{person}{Jeremy Manson}, \bibinfo{person}{William
  Pugh}, {and} \bibinfo{person}{Sarita~V Adve}.}
  \bibinfo{year}{2005}\natexlab{}.
\newblock \showarticletitle{The {Java} memory model}. In
  \bibinfo{booktitle}{\emph{ACM SIGPLAN-SIGACT Symposium on Principles of
  Programming Languages (POPL)}}. \bibinfo{pages}{378--391}.
\newblock


\bibitem[\protect\citeauthoryear{Mantel, Perner, and Sauer}{Mantel
  et~al\mbox{.}}{2014}]%
        {mantel2014noninterference}
\bibfield{author}{\bibinfo{person}{Heiko Mantel}, \bibinfo{person}{Matthias
  Perner}, {and} \bibinfo{person}{Jens Sauer}.}
  \bibinfo{year}{2014}\natexlab{}.
\newblock \showarticletitle{Noninterference under weak memory models}. In
  \bibinfo{booktitle}{\emph{IEEE Computer Security Foundations Symposium
  (CSF)}}. IEEE, \bibinfo{pages}{80--94}.
\newblock


\bibitem[\protect\citeauthoryear{Murray and Lowe}{Murray and Lowe}{2009}]%
        {Murray_Lowe_09}
\bibfield{author}{\bibinfo{person}{Toby Murray} {and} \bibinfo{person}{Gavin
  Lowe}.} \bibinfo{year}{2009}\natexlab{}.
\newblock \showarticletitle{On Refinement-Closed Security Properties and
  Nondeterministic Compositions}. In \bibinfo{booktitle}{\emph{International
  Workshop on Automated Verification of Critical Systems}}
  \emph{(\bibinfo{series}{Electronic Notes in Theoretical Computer Science},
  Vol.~\bibinfo{volume}{250})}. \bibinfo{pages}{49--68}.
\newblock
\urldef\tempurl%
\url{https://doi.org/10.1016/j.entcs.2009.08.017}
\showDOI{\tempurl}


\bibitem[\protect\citeauthoryear{Murray and Lowe}{Murray and Lowe}{2010}]%
        {Murray_Lowe_10}
\bibfield{author}{\bibinfo{person}{Toby Murray} {and} \bibinfo{person}{Gavin
  Lowe}.} \bibinfo{year}{2010}\natexlab{}.
\newblock \showarticletitle{Analysing the Information Flow Properties of
  Object-Capability Patterns}. In \bibinfo{booktitle}{\emph{Formal Aspects of
  Security and Trust}} \emph{(\bibinfo{series}{Lecture Notes in Computer
  Science}, Vol.~\bibinfo{volume}{5983})}. \bibinfo{address}{Eindhoven, The
  Netherlands}, \bibinfo{pages}{81--95}.
\newblock
\urldef\tempurl%
\url{https://doi.org/10.1007/978-3-642-12459-4_7}
\showDOI{\tempurl}


\bibitem[\protect\citeauthoryear{Murray, Sison, and Engelhardt}{Murray
  et~al\mbox{.}}{2018}]%
        {Murray_SE_18}
\bibfield{author}{\bibinfo{person}{Toby Murray}, \bibinfo{person}{Robert
  Sison}, {and} \bibinfo{person}{Kai Engelhardt}.}
  \bibinfo{year}{2018}\natexlab{}.
\newblock \showarticletitle{{COVERN}: {A} Logic for Compositional Verification
  of Information Flow Control}. In \bibinfo{booktitle}{\emph{IEEE European
  Symposium on Security and Privacy (EuroS\&P)}}. \bibinfo{address}{London,
  United Kingdom}.
\newblock


\bibitem[\protect\citeauthoryear{Murray, Sison, Pierzchalski, and
  Rizkallah}{Murray et~al\mbox{.}}{2016}]%
        {Murray_SPR_16}
\bibfield{author}{\bibinfo{person}{Toby Murray}, \bibinfo{person}{Robert
  Sison}, \bibinfo{person}{Edward Pierzchalski}, {and}
  \bibinfo{person}{Christine Rizkallah}.} \bibinfo{year}{2016}\natexlab{}.
\newblock \showarticletitle{Compositional Verification and Refinement of
  Concurrent Value-Dependent Noninterference}. In
  \bibinfo{booktitle}{\emph{IEEE Computer Security Foundations Symposium
  (CSF)}}. \bibinfo{pages}{417--431}.
\newblock


\bibitem[\protect\citeauthoryear{O'Hearn}{O'Hearn}{2004}]%
        {OHearn_04}
\bibfield{author}{\bibinfo{person}{Peter~W O'Hearn}.}
  \bibinfo{year}{2004}\natexlab{}.
\newblock \showarticletitle{Resources, concurrency and local reasoning}. In
  \bibinfo{booktitle}{\emph{International Conference on Concurrency Theory
  (CONCUR)}}. Springer, \bibinfo{pages}{49--67}.
\newblock


\bibitem[\protect\citeauthoryear{Roscoe, Woodcock, and Wulf}{Roscoe
  et~al\mbox{.}}{1994}]%
        {roscoe1994non}
\bibfield{author}{\bibinfo{person}{AW Roscoe}, \bibinfo{person}{JCP Woodcock},
  {and} \bibinfo{person}{Lars Wulf}.} \bibinfo{year}{1994}\natexlab{}.
\newblock \showarticletitle{Non-interference through determinism}. In
  \bibinfo{booktitle}{\emph{European Symposium on Research in Computer Security
  (ESORICS)}}. Springer, \bibinfo{pages}{31--53}.
\newblock


\bibitem[\protect\citeauthoryear{Roscoe}{Roscoe}{1995}]%
        {roscoe1995csp}
\bibfield{author}{\bibinfo{person}{A~William Roscoe}.}
  \bibinfo{year}{1995}\natexlab{}.
\newblock \showarticletitle{{CSP} and determinism in security modelling}. In
  \bibinfo{booktitle}{\emph{IEEE Symposium on Security \& Privacy (S\&P)}}.
  IEEE, \bibinfo{pages}{114--127}.
\newblock


\bibitem[\protect\citeauthoryear{Sarkar, Sewell, Nardelli, Owens, Ridge,
  Braibant, Myreen, and Alglave}{Sarkar et~al\mbox{.}}{2009}]%
        {sarkar2009semantics}
\bibfield{author}{\bibinfo{person}{Susmit Sarkar}, \bibinfo{person}{Peter
  Sewell}, \bibinfo{person}{Francesco~Zappa Nardelli}, \bibinfo{person}{Scott
  Owens}, \bibinfo{person}{Tom Ridge}, \bibinfo{person}{Thomas Braibant},
  \bibinfo{person}{Magnus~O Myreen}, {and} \bibinfo{person}{Jade Alglave}.}
  \bibinfo{year}{2009}\natexlab{}.
\newblock \showarticletitle{The semantics of {x86-CC} multiprocessor machine
  code}. In \bibinfo{booktitle}{\emph{ACM SIGPLAN-SIGACT Symposium on
  Principles of Programming Languages (POPL)}}.
\newblock


\bibitem[\protect\citeauthoryear{Schoepe, Murray, and Sabelfeld}{Schoepe
  et~al\mbox{.}}{2020}]%
        {schoepe2020veronica}
\bibfield{author}{\bibinfo{person}{Daniel Schoepe}, \bibinfo{person}{Toby
  Murray}, {and} \bibinfo{person}{Andrei Sabelfeld}.}
  \bibinfo{year}{2020}\natexlab{}.
\newblock \showarticletitle{{VERONICA}: Expressive and Precise Concurrent
  Information Flow Security}. In \bibinfo{booktitle}{\emph{IEEE Computer
  Security Foundations Symposium (CSF)}}. IEEE, \bibinfo{pages}{79--94}.
\newblock


\bibitem[\protect\citeauthoryear{Sison and Murray}{Sison and Murray}{2019}]%
        {Sison_Murray_19}
\bibfield{author}{\bibinfo{person}{Robert Sison} {and} \bibinfo{person}{Toby
  Murray}.} \bibinfo{year}{2019}\natexlab{}.
\newblock \showarticletitle{Verifying That a Compiler Preserves Concurrent
  Value-Dependent Information-Flow Security}. In
  \bibinfo{booktitle}{\emph{International Conference on Interactive Theorem
  Proving (ITP)}}. \bibinfo{pages}{27:1--27:19}.
\newblock


\bibitem[\protect\citeauthoryear{Smith, Coughlin, and Murray}{Smith
  et~al\mbox{.}}{2019}]%
        {smith2019value}
\bibfield{author}{\bibinfo{person}{Graeme Smith}, \bibinfo{person}{Nicholas
  Coughlin}, {and} \bibinfo{person}{Toby Murray}.}
  \bibinfo{year}{2019}\natexlab{}.
\newblock \showarticletitle{Value-Dependent Information-Flow Security on Weak
  Memory Models}. In \bibinfo{booktitle}{\emph{International Symposium on
  Formal Methods (FM)}}. Springer, \bibinfo{pages}{539--555}.
\newblock


\bibitem[\protect\citeauthoryear{Terauchi}{Terauchi}{2008}]%
        {terauchi2008type}
\bibfield{author}{\bibinfo{person}{Tachio Terauchi}.}
  \bibinfo{year}{2008}\natexlab{}.
\newblock \showarticletitle{A type system for observational determinism}. In
  \bibinfo{booktitle}{\emph{IEEE Computer Security Foundations Symposium
  (CSF)}}. IEEE, \bibinfo{pages}{287--300}.
\newblock


\bibitem[\protect\citeauthoryear{Vafeiadis}{Vafeiadis}{2011}]%
        {vafeiadismfps11}
\bibfield{author}{\bibinfo{person}{Viktor Vafeiadis}.}
  \bibinfo{year}{2011}\natexlab{}.
\newblock \showarticletitle{Concurrent Separation Logic and Operational
  Semantics}. In \bibinfo{booktitle}{\emph{Mathematical Foundations of
  Programming Semantics (MFPS)}}. \bibinfo{pages}{335--351}.
\newblock


\bibitem[\protect\citeauthoryear{Vafeiadis and Narayan}{Vafeiadis and
  Narayan}{2013}]%
        {Vafeiadis_Narayan_13}
\bibfield{author}{\bibinfo{person}{Viktor Vafeiadis} {and}
  \bibinfo{person}{Chinmay Narayan}.} \bibinfo{year}{2013}\natexlab{}.
\newblock \showarticletitle{Relaxed separation logic: A program logic for {C11}
  concurrency}. In \bibinfo{booktitle}{\emph{Conference on Object-Oriented
  Programming Systems, Languages, and Applications (OOPSLA)}}.
  \bibinfo{pages}{867--884}.
\newblock


\bibitem[\protect\citeauthoryear{Vaughan and Millstein}{Vaughan and
  Millstein}{2012}]%
        {vaughan2012secure}
\bibfield{author}{\bibinfo{person}{Jeffrey~A Vaughan} {and}
  \bibinfo{person}{Todd Millstein}.} \bibinfo{year}{2012}\natexlab{}.
\newblock \showarticletitle{Secure information flow for concurrent programs
  under {Total Store Order}}. In \bibinfo{booktitle}{\emph{IEEE Computer
  Security Foundations Symposium (CSF)}}. IEEE, \bibinfo{pages}{19--29}.
\newblock


\bibitem[\protect\citeauthoryear{\v{S}ev\v{c}\'{i}k and
  Sewell}{\v{S}ev\v{c}\'{i}k and Sewell}{2016}]%
        {Sewell:mappings}
\bibfield{author}{\bibinfo{person}{Jaroslav \v{S}ev\v{c}\'{i}k} {and}
  \bibinfo{person}{Peter Sewell}.} \bibinfo{year}{2016}\natexlab{}.
\newblock \bibinfo{title}{C/C++11 mappings to processors}.
\newblock
\newblock
\newblock
\shownote{\url{https://www.cl.cam.ac.uk/~pes20/cpp/cpp0xmappings.html}.
  Accessed 2021-08-12.}


\bibitem[\protect\citeauthoryear{Yan}{Yan}{2021}]%
        {SupplementaryMaterial}
\bibfield{author}{\bibinfo{person}{Pengbo Yan}.}
  \bibinfo{year}{2021}\natexlab{}.
\newblock \bibinfo{booktitle}{\emph{{SecRSL: Security Separation Logic for C11
  Release-Acquire Concurrency - Coq Formalisation}}}.
\newblock
\urldef\tempurl%
\url{https://doi.org/10.5281/zenodo.5493554}
\showDOI{\tempurl}


\bibitem[\protect\citeauthoryear{Yang}{Yang}{2007}]%
        {yang2007relational}
\bibfield{author}{\bibinfo{person}{Hongseok Yang}.}
  \bibinfo{year}{2007}\natexlab{}.
\newblock \showarticletitle{Relational separation logic}.
\newblock \bibinfo{journal}{\emph{Theoretical Computer Science}}
  \bibinfo{volume}{375}, \bibinfo{number}{1-3} (\bibinfo{year}{2007}),
  \bibinfo{pages}{308--334}.
\newblock


\bibitem[\protect\citeauthoryear{Zdancewic and Myers}{Zdancewic and
  Myers}{2003}]%
        {zdancewic2003observational}
\bibfield{author}{\bibinfo{person}{Steve Zdancewic} {and}
  \bibinfo{person}{Andrew~C Myers}.} \bibinfo{year}{2003}\natexlab{}.
\newblock \showarticletitle{Observational determinism for concurrent program
  security}. In \bibinfo{booktitle}{\emph{IEEE Computer Security Foundations
  Workshop (CSFW)}}. IEEE, \bibinfo{pages}{29--43}.
\newblock


\end{thebibliography}

\ifExtended
\newpage
\appendix
\section{Formal Definitions}\label{app}

\subsection{Assertion Semantics}\label{app:assertion}
\begin{definition}[Assertion Semantics]\label{defn:assn-sem}
  Let \setof{-} be a function from assertions to sets of pairs of heaps~$(h,h')$
  defined as follows.\\[-1.5em]
\begin{align*}
	\setof{\Afalse} \defeq&\ \emptyset\\
	\setof{\Aemp} \defeq&\ \{(\emptyset,\emptyset)\}\\
	\setof{\Aimplies{P}{Q}} \defeq&\ \{(h,h')\mid (h,h')\in \setof{P}\implies (h,h')\in \setof{Q}\}\\
	\setof{\Alow{e}{e'}} \defeq&\ \metaITE{e=e'}{\{(\emptyset,\emptyset)\}}{\emptyset}\\
	\setof{\Aforall{x}{x'}{P}} \defeq&\ \{(h,h')\mid \forall (v, v').~(h,h')\in \setof{P\left[(v,v')/(x,x')\right]}\}\\
	\setof{\Astar{P}{Q}} \defeq&\ \{(h1\oplus h2, h1'\oplus h2') \mid (h1,h1')\in \setof{P} \land (h2,h2')\in \setof{Q}\}\\
	\setof{\Aptu{\ell}{\ell'}} \defeq&\ \{(\{\mapsNA{\ell}{\uninit}\},\{\mapsNA{\ell'}{\uninit}\})\}\\
	\setof{\Apt{\ell}{\ell'}{v}{v'}} \defeq&\ \{(\{\mapsNA{\ell}{v}\},\{\mapsNA{\ell'}{v'}\})\}\\
	\setof{\Ainit{\ell}} \defeq&\ \{(h,h)\mid h=\{\mapsA{\ell}{\Ffalse}{\Femp}{\False}{\True}\}\}\\
	\setof{\Arel{\ell}{Q}} \defeq&\ \{(h,h)\mid h=\{\mapsA{\ell}{Q}{\Femp}{\False}{\False}\}\}\\
	\setof{\Aacq{\ell}{Q}} \defeq&\ \{(h,h)\mid h=\{\mapsA{\ell}{\Ffalse}{Q}{\False}{\False}\}\}\\
	\setof{\Armw{\ell}{Q}} \defeq&\ \{(h,h)\mid h=\{\mapsA{\ell}{\Ffalse}{Q}{\True}{\False}\}\}\\
	\text{where}\ \Ffalse\defeq&(\lambda v.~\False)\ and\ \Femp\defeq(\lambda v.~\Aemp).	
\end{align*}
\end{definition}

\subsection{Local Relational Validity}

The definitions of local relational validity makes use of the following
auxiliary functions.

Given an execution $\rchi = \exec{\A}{\lab}{\sbo}{\rf}{\mo}{\sco}$ and a action $a\in \A$, we define:\\
\begin{center}
$\SBin{\chi}{a}\defeq \{(``sb",b,a)\mid \sbo(b,a)\}$\\
$\SBout{\chi}{a}\defeq \{(``sb",a,b)\mid \sbo(a,b)\}$\\
$\SWin{\chi}{a}\defeq \{(``sw",b,a)\mid \rf(b,a)\land \isAcq{\lab(a)}\land \isRel{\lab(b)}\}$\\
$\SWout{\chi}{a}\defeq \{(``sw",a,b)\mid \rf(a,b)\land \isAcq{\lab(b)}\land \isRel{\lab(a)}\}$\\
$\Pre{\chi}{\V}\defeq \{a\mid \exists b\in \V. \sbo(a,b)\lor (\rf(a,b)\land \isAcq{\lab(b)}\land \isRel{\lab(a)})\lor$\\
					$\quad \quad (\rf(a,b)\land (\isNA{\lab(b)}\lor \isNA{\lab(a)}))\}$\\
$\Resp{\chi}{\V}\defeq \bigcup_{a\in\V}(\SBout{\chi}{a}\union \SWin{\chi}{a})$
\end{center}
\ \\
\begin{definition}[Local Relational Validity]\label{defn:local-valid}
Given an execution $\rchi = \exec{\A}{\lab}{\sbo}{\rf}{\mo}{\sco}$,
a set of actions $V\subseteq\A$,
another label function $\lab'$,
two heap maps $(\hmap:\Resp{\chi}{V}\rightarrow \Heap)$ and $(\hmap':\Resp{\chi'}{\V}\rightarrow \Heap)$, and a locations set $\HS$, and letting
$\rchi' = \exec{\A}{\lab'}{\sbo}{\rf}{\mo}{\sco}$, we define
\emph{local relational validity} as follows and denote it
$\Valid{\rchi}{\V}{\lab'}{\hmap}{\hmap'}{\HS}$. It holds if and only if\\
\\
\hspace*{0.3cm} For all a $\in\V$, exists $ \ell, b, v, v', v_1, v_2, Q, Q', Z, h_F, h_F', h_1, h_1',h_{sink}, h_{sink}', $ such that\\
\hspace*{0.7cm}$\left(\begin{array}{l}
	\lab(a) = \skiplab ~\land~ \lab'(a) = \skiplab \\ 
	\land~ \hmap(\SBin{\chi}{a}) =  \hmap(\SBout{\chi}{a}) \oplus h_{sink} \\
	\land~ \hmap'(\SBin{\chi'}{a}) =  \hmap'(\SBout{\chi'}{a}) \oplus h'_{sink}
\end{array}\right)$\\

$\lor \left(\begin{array}{l}
	\lab(a) = \Alab{F}{\ell} ~\land~ \lab'(a) = \Alab{F}{\ell} ~\land~ (F=\high\iff \ell\in \HS) \\ 
	\land~ \hmap(\SBout{\chi}{a}) =  \hmap(\SBin{\chi}{a}) \oplus \{\mapsA{\ell}{Q}{Q}{b}{\False}\} \\
	\land~ \hmap'(\SBout{\chi'}{a}) =  \hmap'(\SBin{\chi'}{a}) \oplus \{\mapsA{\ell}{Q}{Q}{b}{\False}\}
\end{array}\right)$\\

$\lor \left(\begin{array}{l}
	\lab(a) = \Alab{F}{\ell} ~\land~ \lab'(a) = \Alab{F}{\ell} ~\land~ (F=\high\iff \ell\in \HS) \\ 
	\land~ \hmap(\SBout{\chi}{a}) =  \hmap(\SBin{\chi}{a}) \oplus \{\mapsNA{\ell}{\uninit}\} \\
	\land~ \hmap'(\SBout{\chi'}{a}) =  \hmap'(\SBin{\chi'}{a}) \oplus \{\mapsNA{\ell}{\uninit}\}
\end{array}\right)$\\

$\lor \left(\begin{array}{l}
	\lab(a) = \Wlab{\na}{\ell}{v} ~\land~ \lab'(a) = \Wlab{\na}{\ell}{v'} ~\land~ (v\neq v' \implies \ell \in \HS)\\ 
	\land~ \hmap(\SBin{\chi}{a})(\ell) \in \{\NA{\uninit},~\NA{\_} \} \\
	\land~ \hmap'(\SBin{\chi'}{a})(\ell) \in \{\NA{\uninit},\NA{\_} \} \\
	\land~ \hmap(\SBout{\chi}{a}) =  \hmap(\SBin{\chi}{a}) [\mapsNA{\ell}{v}]\\
	\land~ \hmap'(\SBout{\chi'}{a}) =  \hmap'(\SBin{\chi'}{a}) [\mapsNA{\ell}{v'}]
\end{array}\right)$\\

$\lor \left(\begin{array}{l}
	\lab(a) = \Rlab{\na}{\ell}{v} ~\land~ \lab'(a) = \Rlab{\na}{\ell}{v'}\\
	\land~ \hmap(\SBin{\chi}{a})(\ell) = \NA{\_} 
	\land~ \hmap'(\SBin{\chi'}{a})(\ell) = \NA{\_} \\
	\land~ \hmap(\SBin{\chi}{a}) =  \hmap(\SBout{\chi}{a})
	\land~ \hmap'(\SBin{\chi'}{a}) =  \hmap'(\SBout{\chi'}{a})
\end{array}\right)$\\

$\lor \left(\begin{array}{l}
	\lab(a) = \lab'(a) = \Rlab{Z}{\ell}{v} ~\land~ Z\in\{\rlx,\acq,\scx\} \\ 
	\land~ (\hmap(\SWin{\chi}{a}),\hmap'(\SWin{\chi'}{a}))\in \setof{Q(v)} \land \Precise{Q(v)} \\
	\land~ \hmap(\SBin{\chi}{a}) = \{\mapsA{\ell}{\Ffalse}{Q}{\False}{\True}\} \oplus h_F\\
	\land~ \hmap'(\SBin{\chi'}{a}) = \{\mapsA{\ell}{\Ffalse}{Q}{\False}{\True}\} \oplus h'_F\\
	\land~ \hmap(\SBout{\chi}{a}) = \hmap(\SWin{\chi}{a}) \oplus h_F \oplus h_1\\
	\land~ \hmap'(\SBout{\chi'}{a}) = \hmap'(\SWin{\chi'}{a}) \oplus h_F' \oplus h_1\\
	\land~ h_1 = \{\mapsA{\ell}{\Ffalse}{Q[v:=\Aemp]}{\False}{\True}\}
\end{array}\right)$\\

$\lor \left(\begin{array}{l}
	\lab(a) = \lab'(a) = \Wlab{Z}{\ell}{v} ~\land~ Z\in\{\rlx,\rel,\scx\} \\ 
	\land~ h_1 = \hmap(\SWout{\chi}{a}) \oplus h_{sink}
	~\land~ h'_1 = \hmap'(\SWout{\chi'}{a}) \oplus h'_{sink}\\
	\land~ \hmap(\SBin{\chi}{a}) = \{\mapsA{\ell}{Q}{\Femp}{\False}{b}\} \oplus h_1 \oplus h_F\\
	\land~ \hmap'(\SBin{\chi'}{a}) = \{\mapsA{\ell}{Q}{\Femp}{\False}{b}\} \oplus h'_1 \oplus h'_F\\
	\land~ \hmap(\SBout{\chi}{a}) = \{\mapsA{\ell}{Q}{\Femp}{\False}{\True}\} \oplus h_F\\
	\land~ \hmap'(\SBout{\chi'}{a}) = \{\mapsA{\ell}{Q}{\Femp}{\False}{\True}\} \oplus h'_F\\
	\land~ (h_1,h'_1)\in\setof{Q(v)} ~\land~ (Z ~=~ \rlx \implies Q(v) = \Aemp)
\end{array}\right)$\\

$\lor \left(\begin{array}{l}
	\lab(a) = \lab'(a) = \RMWlab{Z}{\ell}{v_1}{v_2} ~\land~ Z\ne \na \\ 
	\land~ \hmap(\SBin{\chi}{a})(\ell) = \hmap'(\SBin{\chi'}{a})(\ell) = \hA{\_}{Q}{\True}{\True}\\
	\land~ \hmap(\SBin{\chi}{a}) \oplus \hmap(\SWin{\chi}{a}) = \{\mapsA{\ell}{Q'}{\Femp}{\False}{\False}\}\\
	~~~~~~~~~~~~~~~~~~~~~~~~~~~~~~~ \oplus \hmap(\SBout{\chi}{a}) \oplus \hmap(\SWout{\chi}{a}) \oplus h_{sink}\\
	\land~ \hmap'(\SBin{\chi'}{a}) \oplus \hmap'(\SWin{\chi'}{a}) = \{\mapsA{\ell}{Q'}{\Femp}{\False}{\False}\}\\
	~~~~~~~~~~~~~~~~~~~~~~~~~~~~~~~ \oplus \hmap'(\SBout{\chi'}{a}) \oplus \hmap'(\SWout{\chi'}{a}) \oplus h'_{sink}\\
	\land~ (\hmap(\SWin{\chi}{a}),\hmap'(\SWin{\chi'}{a}))\in \setof{Q(v)}\\
	\land~ (\hmap(\SWout{\chi}{a}) \oplus h_{sink},\hmap'(\SWout{\chi'}{a}) \oplus h'_{sink})\in\setof{Q'(v')}\\
	\land~ (Z \in\{\rlx,\rel\} \implies Q(v) = \Aemp) ~\land~ (Z \in\{\rlx,\acq\} \implies Q'(v') = \Aemp)
\end{array}\right)$\\

where ``\_'' denotes a wildcard here and in the following pages.
\end{definition}

\subsection{Partial Consistency}

The definition of partial consistency refers to various axioms of the
memory model, introduced in~\citet{Vafeiadis_Narayan_13}.

\begin{definition}[Partial Consistency]\label{defn:pcons}
	Given a set $\V \subseteq \AName$ and three binary relations~$\sbo$, $\rf$, $\mo$ on $\V$. a subset $\V' \subseteq \V$, and an action labelling
	function~$lab'$ whose domain is~$\V'$, we define \emph{partial consistency}
	as follows, and denote it~$\Pcons{\V}{\V'}{\lab'}{\sbo}{\rf}{\mo}$:\\[1ex]
	$\Pcons{\V}{\V'}{\lab'}{\sbo}{\rf}{\mo} \defeq$\\
	$\hspace*{1cm}\mathsf{IrreflexiveHB}(\lab',\sb,\rf,\mo) \land \mathsf{ConsistentAlloc}(\lab') \ \land$\\
	$\hspace*{1cm}(\forall a, (a\in V' \lor \lab'(a) = \skiplab))\ \land$\\
	$\hspace*{1cm}(\forall a\in \V' \land \rf(a)\in \V',\ \exists\ \ell, v.\ \lab'(a) = \Rlab{\_}{\ell}{v} \land \lab'(\rf(a)) = \Wlab{\_}{\ell}{v})$\\
\end{definition}

\subsection{The Meaning of SecRSL Judgements}
The meaning of SecRSL judgements makes use of the concept of
configuration safety, which we define formally as follows. It is
captured by the inductive predicate~$\mathsf{safe}$.

\begin{definition}[Configuration Safety]\label{defn:safe}
	Given two sets of actions $\A_{ctx}$ and $\A_{prg}$, an execution $\rchi = \exec{\A_{ctx}\uplus \A_{prg}}{\lab}{\sbo}{\rf}{\mo}{\sco}$,
	a natural number $n\in\nat$, a value $\res$, a set of actions $V$, another label function $\lab'$, two heap maps $(\hmap:\Resp{\chi}{V}\rightarrow \Heap)$ and $(\hmap':\Resp{\chi'}{\V}\rightarrow \Heap)$, a location set $\HS$, the start action $\fst\in\A_{prg}$ and the final action $\lst\in\A_{prg}$, an expression $E'$, a function $Q$ from a pair of (return) values to a set of pairs of heaps, and a function $HQ$ from a pair of values to a location set, then,
	$\simsafe{n}{\V}{lab'}{\hmap}{\hmap'}{\HS}$ is defined by recursion on $n$:\\
	\\
	\hspace*{0.3cm} $ \simsafe{0}{\V}{lab'}{\hmap}{\hmap'}{\HS}$ always hold. \\
	\\
	\hspace*{0.3cm} $ \simsafe{n+1}{\V}{lab'}{\hmap}{\hmap'}{\HS} $ holds if and only if the following conditions all hold:\\
	\\
	\hspace*{0.3cm} $\bullet $ If $ \lst \in \V,$ Exists $\res'. ~\exec{\res'}{\A_{prg}}{\lab'}{\sbo}{\fst}{\lst} \in \setof{E'} ~ \land $\\
	\hspace*{0.6cm} $(\hmap(\SBout{\chi}{\lst}), ~ \hmap'(\SBout{\chi'}{\lst})) \in Q(\res,\res') \land HQ(\res,\res')\subseteq \HS$;\\
	\\
	\hspace*{0.3cm} $\bullet $ For all $ a\in \A_{prg}\setminus \V $ such that $ \Pre{\chi}{\{a\}} \subseteq \V,$ Exists $ action',\lab2',\hmap2,\hmap2',\HS2.$ \\
	\hspace*{0.6cm} $(\forall x, (x=a \land \lab2'(x) = actions')\lor (x\ne a \land \lab2'(x) =\lab'(x)) ) ~\land~$\\
	\hspace*{0.6cm} $\Pcons{\A_{ctx}\uplus \A_{prg}}{\V\union\{a\}}{\lab2'}{\sbo}{\rf}{\mo} ~\land~\Pre{\chi'}{\{a\}} \subseteq \V~\land$\\
	\hspace*{0.6cm} $\Valid{\rchi}{\V\union\{a\}}{\lab2'}{\hmap \union \hmap2}{\hmap' \union \hmap2'}{\HS2}~\land~$\\
	\hspace*{0.6cm} $(\forall \ell\in \HS2, \exists m.\ \lab2'(m) = \Alab{\_}{\ell}) \land$\\
	\hspace*{0.6cm} $\simsafe{n}{\V\union\{a\}}{\lab2'}{\hmap \union \hmap2}{\hmap' \union \hmap2'}{\HS2}$\\
	\\
	\hspace*{0.3cm} $\bullet $ For all $ a\in \A_{ctx}\setminus \V $ such that $ \Pre{\chi}{\{a\}} \subseteq \V,$ \\
	\hspace*{0.6cm} For all $ action',\lab2',\hmap2,\hmap2',\HS2\textbf{ such that}$ \\
	\hspace*{0.6cm} $(\forall x, (x=a \land \lab2'(x) = actions')\lor (x\ne a \land \lab2'(x) =\lab'(x)) ) ~\land~$\\
	\hspace*{0.6cm} $\Pcons{\A_{ctx}\uplus \A_{prg}}{\V\union\{a\}}{\lab2'}{\sbo}{\rf}{\mo} ~\land~\Pre{\chi'}{\{a\}} \subseteq \V~\land$\\
	\hspace*{0.6cm} $\Valid{\rchi}{\V\union\{a\}}{\lab2'}{\hmap \union \hmap2}{\hmap' \union \hmap2'}{\HS2}~\land~$\\
	\hspace*{0.6cm} $(\forall \ell\in \HS2, \exists m.\ \lab2'(m) = \Alab{\_}{\ell}),$\textbf{ then} \\
	\hspace*{0.6cm} $\simsafe{n}{\V\union\{a\}}{\lab2'}{\hmap \union \hmap2}{\hmap' \union \hmap2'}{\HS2}$\\
\end{definition}

With configuration safety defined, we can formally define relational validity,
i.e.\ the semantic meaning of SecRSL judgements.

\begin{definition}[Relational Validity]\label{defn:triple}
	The SecRSL judgement~$\Triple{HP}{P}{E}{E'}{(y,y').Q}{HQ}$ holds if and only if \\
	\hspace*{0.5cm} \textbf{For all} $n \in \nat, R \in \Assn, \A_{ctx} $ \textbf{such that} $\A_{ctx}\inter \A_{prg} = \emptyset$, \\
	\hspace*{1.6cm} $\sbo,\rf,\mo,\sco,\fst,\lst,\lab',\hmap,\hmap',\HS,\lab_{ctx},\lab_{prg},\lab = \lab_{prg}\union \lab_{ctx},$\\
	\hspace*{1.6cm} $\rchi = \exec{\A_{ctx}\uplus \A_{prg}}{\lab}{\sbo}{\rf}{\mo}{\sco}, \rchi' =\exec{\A_{ctx}\uplus \A_{prg}}{\lab'}{\sbo}{\rf}{\mo}{\sco},$\\
	\hspace*{0.1cm} \textbf{such that} $ \exec{\res}{\A_{prg}}{\lab_{prg}}{\sbo\inter(\A_{prg}\times \A_{prg})}{\fst}{\lst} \in \setof{E} ~ \land $ \\
	\hspace*{1.7cm} $\Consistent{\rchi} \land \Pcons{\A_{ctx}\uplus \A_{prg}}{\V}{\lab'}{\sbo}{\rf}{\mo}~\land$ \\
	\hspace*{1.7cm} $ (\exists! a.~\sbo(a,\fst)) ~ \land ~ (\exists! b.~\sbo(\lst,b)) ~\land~$ \\
	\hspace*{1.7cm} $\V \subseteq \A_{ctx} \land \Pre{\chi}{V} \land \Pre{\chi'}{V} \subseteq \V \land a\in \V ~\land$ \\
	\hspace*{1.7cm} $\Valid{\rchi}{\V}{\lab'}{\hmap}{\hmap'}{\HS}~\land~ (\forall \ell\in \HS, \exists m.\ \lab'(m) = \Alab{\_}{\ell})~\land$\\
	\hspace*{1.7cm} $ (\hmap (\SBout{\chi}{a}), \hmap' (\SBout{\chi'}{a})) \in \setof{P*R} ~\land HP \subseteq \HS,$ \textbf{then}\\
	\hspace*{0.6cm} $\mathsf{safe}^n_\chi (\res,\V,\lab',\hmap,\hmap',\HS,\A_{prg},\A_{ctx},\fst,\lst,E',\setof{Q*R},HQ), $\\
\end{definition}

\section{Proof Sketches for Case Studies}\label{app:proofs}

\subsection{The Spinlock}\label{app:proofs-spinlock}

\begin{center}
	$\PQ{\textbf{Let}\ \ Q_J(v)\defeq (v=0\land \Aemp)\lor (v=1\land J)}$\\
	$\PQ{\Lock{x}{J} \defeq \Arel{x}{Q_J}\star \Armw{x}{Q_J}\star \Ainit{x}}$\\
	\begin{tabular}{l@{\quad}|@{\quad}l}
		$new\_lock()\defeq$& $lock(x)\defeq$\\
		$\ \ \ \ \PQ{\{J\}}$& $\PQ{\ \ \ \ \{\Lock{x}{J}\}}$\\
		$\ \ \ \ \Let{x}{\Alloc{\low}}{}$& $\ \ \ \ \mathbf{repeat}$\\
		$\ \ \ \ \PQ{\{J\star \Arel{x}{Q_J}\star}$& $\PQ{\ \ \ \ \ \ \ \ \{\Lock{x}{J}\}}$\\
		$\PQ{\ \ \ \ \ \ \Armw{x}{Q_J}\}}$& $\ \ \ \ \ \ \ \ spin(x);$\\
		$\ \ \ \ \Store{x}{\rel}{1}$& $\PQ{\ \ \ \ \ \ \ \ \{\Lock{x}{J}\}}$\\
		$\PQ{\ \ \ \ \{\Lock{x}{J}\}}$& $\ \ \ \ \ \ \ \ \CAS{\acq}{\rlx}{x}{1}{0}$\\
		$unlock()\defeq$& $\PQ{\ \ \ \ \ \ \ \ \{(y,y').\ \Lock{x}{J}}$\\
		$\PQ{\ \ \ \ \{J\star \Lock{x}{J}\}}$& $\PQ{\ \ \ \ \ \ \ \ \ \ \ \star\ \  Q_J(y)\}}$\\
		$\ \ \ \ \Store{x}{\rel}{1}$& $\ \ \ \ \mathbf{end}$\\
		$\PQ{\ \ \ \ \{\Ainit{x}\star \Lock{x}{J}\}}$& $\PQ{\ \ \ \ \{J\star \Lock{x}{J}\}}$\\
		$\PQ{\ \ \ \ \{\Lock{x}{J}\}}$&
	\end{tabular}\\
	\captionof{figure}{Verification of the spinlock module.}
	\label{spinlock}
\end{center}\par

\subsection{Mixed-Sensitivity Mutex}\label{app:proofs-spinlock-clas}

We define the invariant $LQ$ and predicate $\LockC{\,}$ as follows.

\begin{center}
	$\PQ{LQ \defeq\ \lambda v.\ \metaITE{v=0}{Aemp}{\metaITE{v=1}{\CLow{b}}{\CHigh{b}}}}$\\
	$\PQ{\LockC{a} \defeq \Arel{x}{LQ}\star \Armw{x}{LQ}\star \Ainit{x}}$\\
\end{center}

The the proof for $\newMSM$ is as follows:\\
\begin{tabular}{l}
	$\PQ{\preC{}{\Aemp}}$\\
	$\qquad\Let{b}{\Alloc{\high}}{}$\\
	$\PQ{\postC{b}{\Aptu{b}{b}}}$\\
	$\qquad\Let{a}{\Alloc{\low}}{}$\\
	$\PQ{\postC{b}{\Aptu{b}{b}\sep\Arel{a}{LQ}\sep\Armw{a}{LQ}}}$\\
	$\qquad\Store{b}{\na}{0};$\\
	$\PQ{\postC{b}{\Apt{b}{b}{0}{0}\sep\Arel{a}{LQ}\sep\Armw{a}{LQ}}}$\\
	$\qquad\Store{a}{\rel}{1};$\\
	$\PQ{\postC{b}{\LockC{a}}}$\\
\end{tabular}\\

The proof for $\lockMSM{a}$ is sketched below:\\
\begin{tabular}{l}
	$\PQ{\preC{b}{\LockC{a}}}$\\
	$\mathbf{repeat}$\\
	$\quad\Let{x}{(\Repeat{\Load{a}{\rlx}})}{}$\\
	$\quad\PQ{\postC{b}{(x,x).\ \LockC{a}\land x>0}}$\\
	$\quad\Let{y}{\CAS{\acq}{\rlx}{a}{x}{0}}{}$\\
	$\quad\PQ{\postC{b}{(y,y).\ \LockC{a}\sep (\metaITE{x=y}{LQ(y)}{\Aemp})\land x>0}}$\\
	$\quad \If{(x==y)}{x}{0}$\\
	$\quad\PQ{\postC{b}{(z,z).\ \LockC{a}\sep (\metaITE{x=y}{LQ(y)}{\Aemp})\land ((z>0\land x=y=z)\lor(z=0))}}$\\
	$\mathbf{end}$\\
	$\PQ{\postC{b}{(z,z).\ \LockC{a}\sep (\metaITE{x=y}{LQ(y)}{\Aemp})\land x=y=z}\qquad\implies}$\\
	$\PQ{\postC{b}{(z,z).\ \LockC{a}\sep LQ(z)\land z>0}}$\\
\end{tabular}\\

The other specifications are straightforward rule applications
(e.g.\ can be deduced by applying the split rules in \cref{ASRules} or
the \textsc{Rel-W} rule from \cref{ARules}).

\subsection{Release/Acquire Synchronous Channel}\label{app:syncchannel-proof}

Given an invariant $Q$, we define several auxiliary predicates that we
will use to split $\Aacq{\,}{\,}$ permissions:\\
\hspace*{0.6cm} $Q_{(n)} \defeq\ \lambda v.\ \metaITE{v=n}{Q(v)}{\Aemp}$\\
\hspace*{0.6cm} $Q_{(n,m)} \defeq\ \lambda v.\ \metaITE{v=n \ \mathit{or}\  v=m}{Q(v)}{\Aemp}$\\

Then we define various invariants, where $v$ refers to the value
stored in atomic location~$a$. $QQ$ is the invariant
for the $\Arel{a}{\,}$ permission of both threads, while $SQ$ and $RQ$ are
for the $\Aacq{a}{\,}$ permission of the sender and receiver 
respectively:\\
\hspace*{0.6cm} $QQ \defeq\ \lambda v.\ \metaITE{v=0}{\Aemp}{\metaITE{\Mod{v}{3}=1}{\CLow{b}}{\CHigh{b}}}$\\
\hspace*{0.6cm} $SQ^n \defeq\ \lambda v.\ \metaITE{v>n \ \mathit{and}\  \Mod{v}{3}=0}{\CHigh{b}}{\Aemp}$\\
\hspace*{0.6cm} $RQ^n \defeq\ \lambda v.\ \metaITE{v>n\ \mathit{and}\  \Mod{v}{3}=1}{\CLow{b}}{}$\\
\hspace*{2.35cm} $\metaITE{v>n \ \mathit{and}\  \Mod{v}{3}=2}{\CHigh{b}}{\Aemp}$\\

With split rules (\cref{ASRules}) we can derive: for all $a$ and for all $n$ such that $\Mod{n}{3}=0$,\\
\hspace*{0.6cm} $\Arel{a}{QQ}\iff \Arel{a}{QQ} \sep \Arel{a}{QQ}$\\
\hspace*{0.6cm} $\Aacq{a}{QQ}\iff \Aacq{a}{SQ^0} \sep \Aacq{a}{RQ^0}$\\
\hspace*{0.6cm} $\Aacq{a}{SQ^n}\iff \Aacq{a}{SQ^n_{(n+3)}} \sep \Aacq{a}{SQ^{n+3}}$\\
\hspace*{0.6cm} $\Aacq{a}{RQ^n}\iff \Aacq{a}{RQ^n_{(n+1,n+2)}} \sep \Aacq{a}{RQ^{n+3}}$\\

Finally we define the permissions to send and receive, mentioned in
\cref{sec:syncchannel-ra}:\\
\hspace*{0.6cm} $\Sender{n} \defeq\ \Apt{b}{b}{n}{n}\ \sep\ \Arel{a}{QQ}\ \sep\ \Aacq{a}{SQ^n}\ \sep\ \Ainit{a}$\\
\hspace*{0.6cm} $\Recver{n} \defeq\ \Apt{c}{c}{n}{n}\ \sep\ \Arel{a}{QQ}\ \sep\ \Aacq{a}{RQ^n}\ \sep\ \Ainit{a}$\\

Then we sketch the proofs for each of the three functions. We begin
with $\newchannel$:\\
\begin{tabular}{l}
	$\PQ{\preC{}{\Aemp}}$\\
	$\qquad \Let{b}{\Alloc{\high}}{}$\\
	$\PQ{\postC{b}{\Aptu{b}{b}}}$\\
	$\qquad \Let{c}{\Alloc{\low}}{}$\\
	$\PQ{\postC{b}{\Aptu{b}{b}\ \sep \ \Aptu{c}{c}}}$\\
	$\qquad \Let{a}{\Alloc{\low}}{}$\\
	$\PQ{\postC{b}{\Aptu{b}{b}\ \sep \ \Aptu{c}{c}\ \sep \Arel{a}{QQ}\ \sep\ \Aacq{a}{QQ}}}$\\
	$\qquad \Store{b}{\na}{0};$\\
	$\PQ{\postC{b}{\Apt{b}{b}{0}{0}\ \sep \ \Aptu{c}{c}\ \sep \Arel{a}{QQ}\ \sep\ \Aacq{a}{QQ}}}$\\
	$\qquad \Store{c}{\na}{0};$\\
	$\PQ{\postC{b}{\Apt{b}{b}{0}{0}\ \sep \ \Apt{c}{c}{0}{0}\ \sep \Arel{a}{QQ}\ \sep\ \Aacq{a}{QQ}}}$\\
	$\qquad \Store{a}{\rel}{0};$\\
	$\PQ{\postC{b}{\Apt{b}{b}{0}{0}\ \sep \ \Apt{c}{c}{0}{0}\ \sep \Arel{a}{QQ}\ \sep\ \Aacq{a}{QQ}\ \sep\ \Ainit{a}}}$\\
	$\qquad \PQ{\iff}\qquad \PQ{\postC{b}{\Sender{0}\ \sep\ \Recver{0}}}$\\
\end{tabular}
\ \\

The proof for the two executions of $\send{v}{\ishigh}$ and $\send{v'}{\ishigh}$ is sketched thus:\\

Let $HL = (\metaITE{\ishigh}{\Aemp}{\Alow{v}{v'}}),$\\
\begin{tabular}{l}
	$\PQ{\preC{b}{\Sender{n}\ \sep\ HL}}\qquad \PQ{\iff}$\\
	$\PQ{\preC{b}{\Apt{b}{b}{n}{n}\sep\Arel{a}{QQ}\sep\Aacq{a}{SQ^n}\sep\Ainit{a}\sep HL}}$\\
	$\qquad \Let{x}{\Load{b}{\na}}{}$\\
	$\PQ{\postC{b}{\Apt{b}{b}{n}{n}\sep\Arel{a}{QQ}\sep\Aacq{a}{SQ^n}\sep\Ainit{a}\sep HL\land x=n}}$\\
	$\qquad \Store{b}{\na}{v$ (or $v'$ in the other execution)$};$\\
	$\PQ{\postC{b}{\Apt{b}{b}{v}{v'}\sep\Arel{a}{QQ}\sep\Aacq{a}{SQ^n}\sep\Ainit{a}\sep HL\land x=n}}$\\
	$\qquad \Store{a}{\rel}{x+(\ishigh\ ?\ 2:1)};$\\
	$\PQ{\postC{b}{\Arel{a}{QQ}\sep\Aacq{a}{SQ^n}\sep\Ainit{a}\sep\Ainit{a}\land x=n}}\ \PQ{\iff}$ (by the split rules above)\\
	$\PQ{\postC{b}{\Arel{a}{QQ}\sep\rPQ{\Aacq{a}{SQ^n_{(n+3)}}}\sep\Aacq{a}{SQ^{n+3}}\sep\rPQ{\Ainit{a}}\sep\Ainit{a}\land x=n}}$\\
	$\rPQ{\mathsf{(using\ the\ Frame\ Rule)}}$\\
	$\qquad \mathbf{repeat}$\\
	$\quad \rPQ{\postC{b}{\Aacq{a}{SQ^n_{(n+3)}}\sep\Ainit{a}\land x=n}}$\\
	$\qquad \quad \Let{z}{\Load{a}{\acq}}{}$\\
	$\quad \rPQ{\postC{b}{(z,z).\ \ SQ^n_{(n+3)}(z) \sep \Aacq{a}{SQ^n_{(n+3)}\left[z:=\Aemp\right]}\land x=n}}$\\
	$\qquad \quad \quad \If{z==x+3}{1}{0}$\\
	$\rPQ{\postC{b}{(y,y).\ \ SQ^n_{(n+3)}(z) \sep \Aacq{a}{SQ^n_{(n+3)}\left[z:=\Aemp\right]}\land x=n \land (y>0\ ?\ (z=n+3) : (z=0))}}$\\
	$\qquad \mathbf{end}$\\
	$\quad \rPQ{\postC{b}{SQ^n_{(n+3)}(n+3) \sep \Aacq{a}{SQ^n_{(n+3)}\left[n+3:=\Aemp\right]}\land x=n}}\quad \qquad \rPQ{\iff}$\\
	$\quad \rPQ{\postC{b}{\CHigh{b} \sep \Aacq{a}{(\lambda v.\ \Aemp)}\land x=n}}$\\
	$\rPQ{\mathsf{(recover\ Frames)}}$\\
	$\PQ{\postC{b}{\Arel{a}{QQ}\sep\rPQ{\Aacq{a}{(\lambda v.\ \Aemp)}}\sep\Aacq{a}{SQ^{n+3}}\sep\rPQ{\CHigh{b}}\sep\Ainit{a}\land x=n}}$\\
	$\qquad \Store{b}{\na}{x+3};$\\
	$\PQ{\postC{b}{\Arel{a}{QQ}\sep\Aacq{a}{(\lambda v.\ \Aemp)}\sep\Aacq{a}{SQ^{n+3}}\sep\Apt{b}{b}{n+3}{n+3}\sep\Ainit{a}}}$\\
	$\qquad \PQ{\iff}\qquad \PQ{\postC{b}{\Sender{n+3}}}$\\
\end{tabular}\\
\ \\
The main idea of this proof is to split the Acq-permission and use the
Frame rule to
cut one part of the permission (as shown by the two colours). For each round of the two-way transfer,
we use only the part of the Acq-permission which is for that particular
round and keep the
remaining permissions for future rounds.\\

The verification of $\recv{\,}$ is a little more complicated.
The Acq-permission of a
certain round $n$ has two possible values: $n+1$ and $n+2$, while which one will be
read is uncertain. Thus after $\recv{\,}$, one of them will be consumed while
the other one
will remain: the later one is the ``\ldots'' in the postcondition for
$\recv{\,}$ from \cref{sec:syncchannel-ra}. Thus the sketch of the proof for
$\recv{d}$:

\begin{tabular}{l}
	$\PQ{\preC{b,d}{\Recver{n}\ \sep\ \CHigh{d}}}\qquad \PQ{\iff}$\\
	$\PQ{\preC{b,d}{\Apt{c}{c}{n}{n}\sep\Arel{a}{QQ}\sep\Aacq{a}{RQ^n}\sep\Ainit{a}\sep\CHigh{d}}}$\\
	$\qquad \Let{t}{\Load{c}{\na}}{}$\\
	$\PQ{\postC{b,d}{\Apt{c}{c}{n}{n}\sep\Arel{a}{QQ}\sep\Aacq{a}{RQ^n}\sep\Ainit{a}\sep\CHigh{d}\land t=n}\ \PQ{\iff}}$\\
	$\PQ{\{\Apt{c}{c}{n}{n}\sep\Arel{a}{QQ}\sep\rPQ{\Aacq{a}{RQ^n_{(n+1,n+2)}}}\sep\Aacq{a}{RQ^{n+3}}\sep}$\\
	$\qquad \qquad \qquad \qquad \qquad \qquad \qquad \qquad \qquad \qquad\PQ{\rPQ{\Ainit{a}}\sep\CHigh{d}\land t=n\},\ [b,d]}$\\
	$\rPQ{\mathsf{(use\ Frame\ Rule)}}$\\
	$\qquad \mathbf{let}\ lv\ =\ \mathbf{(repeat}$\\
	$\rPQ{\postC{b,d}{\Aacq{a}{RQ^n_{(n+1,n+2)}}\sep\Ainit{a}\land t=n}}$\\
	$\qquad \qquad \qquad \qquad \Let{z}{\Load{a}{\acq}}{}$\\
	$\rPQ{\postC{b,d}{(z,z).\ \ RQ^n_{(n+3)}(z) \sep \Aacq{a}{RQ^n_{(n+1,n+2)}\left[z:=\Aemp\right]}\land t=n}}$\\
	$\qquad \qquad \qquad \qquad \quad \If{(z==t+1)||(z==t+2)}{z}{0}$\\
	$\rPQ{\{(y,y).\ \ RQ^n_{(n+3)}(z) \sep \Aacq{a}{RQ^n_{(n+1,n+2)}\left[z:=\Aemp\right]}}$\\
	$\qquad \qquad \qquad \quad \qquad \quad\rPQ{\land t=n\land (y=0\lor y=z=n+1\lor y=z=n+2)\},\ [b,d]}$\\
	$\qquad \qquad \qquad \quad \ \mathbf{end)}\ \mathbf{in}$\\
	$\rPQ{\postSC{b,d}{(lv,lv).\ \ RQ^n_{(n+3)}(lv) \sep \Aacq{a}{RQ^n_{(n+1,n+2)}\left[lv:=\Aemp\right]}\land t=n\land (lv=n+1\lor lv=n+2)}}$\\
	$\rPQ{\implies}\rPQ{\{(\metaITE{lv=n+1}{\CHigh{b}}{\CLow{b}}) \sep \Aacq{a}{RQ^n_{(2n+3-lv)}}\sep}$\\
		$\qquad\qquad\qquad\rPQ{\land t=n\land (lv=n+1\lor lv=n+2)\},\left[b,d\right]}$\\
	$\rPQ{\mathsf{(recover\ Frames)}}$\\
	$\rPQ{\PQ{\{}(\metaITE{lv=n+1}{\CHigh{b}}{\CLow{b}}) \sep \Aacq{a}{RQ^n_{(2n+3-lv)}}\sep\PQ{\Apt{c}{c}{n}{n}\sep}}$\\
		$\qquad\PQ{\Arel{a}{QQ}\sep\Aacq{a}{RQ^{n+3}}\sep\CHigh{d}\land t=n\land (lv=n+1\lor lv=n+2)\},\left[b,d\right]}$\\
	$\qquad \Let{v}{\Load{b}{\na}}{}$\\
	$\qquad \Store{d}{\na}{v};$\\
	$\PQ{\{(\metaITE{lv=n+1}{\CHigh{d}\sep\CHigh{b}}{\CLow{d}\sep\CLow{b}})\sep}$\\
	$\qquad\PQ{\Apt{c}{c}{n}{n}\sep\Aacq{a}{RQ^n_{(2n+3-lv)}} \sep\Arel{a}{QQ}\sep\Aacq{a}{RQ^{n+3}}}$\\
	$\qquad\PQ{\land t=n\land (lv=n+1\lor lv=n+2)\},\left[b,d\right]}$\\
	$\qquad \Store{a}{\rel}{t+3};$\\
	$\PQ{\{(\metaITE{lv=n+1}{\CHigh{d}}{\CLow{d}})\sep\Apt{c}{c}{n}{n}\sep\Aacq{a}{RQ^n_{(2n+3-lv)}}\sep}$\\
	$\qquad\PQ{\Arel{a}{QQ}\sep\Aacq{a}{RQ^{n+3}}\land t=n\land (lv=n+1\lor lv=n+2)\},\left[b,d\right]}$\\
	$\qquad \Store{c}{\na}{t+3};$\\
	$\PQ{\{(\metaITE{lv=n+1}{\CHigh{d}}{\CLow{d}})\sep\Recver{n+3}\sep\Aacq{a}{RQ^n_{(2n+3-lv)}}}$\\
	$\qquad\PQ{\land t=n\land (lv=n+1\lor lv=n+2)\},\left[b,d\right]}$\\
	$\qquad lv-t;$\\
	$\PQ{\{(y,y).\ (\metaITE{y=1}{\CHigh{d}}{\CLow{d}})\sep\Recver{n+3}\sep\Aacq{a}{RQ^n_{(2n+3-lv)}}}$\\
	$\qquad\PQ{\land (y=1\lor y=2)\},\left[b,d\right]}$\\
\end{tabular}

\newpage

  \section{Sequentially-Consistent Channel Implementation}\label{app:baseline}

A sequentially-consistent implementation of the synchronous channel, used
as a baseline in the evaluation from \cref{sec:eval}, is shown in \cref{fig:baseline}.
This implementation was produced by
taking the release/acquire channel from \cref{fig:syncchannel}
and marking all accesses to shared variables with the sequentially-consistent
mode~$\scx$.

\begin{figure}[h!]
	\begin{minipage}{0.9\textwidth}
		\begin{tabular}{ll}
			\begin{minipage}{0.4\textwidth}
				$\newchannel\ \ \defeq\ \ $ \\
				\ \ \ \ \begin{tabular}{l}
					$\Let{b}{\Alloc{}}{}$\\
					$\Let{c}{\Alloc{}}{}$\\
					$\Let{a}{\Alloc{}}{}$\\
					$\Store{b}{\scx}{0};$\\
					$\Store{c}{\scx}{0};$\\
					$\Store{a}{\scx}{0};$\\
				\end{tabular}
			\end{minipage} &
			\begin{minipage}{0.4\textwidth}
				$\send{v}{\ishigh}\ \ \defeq$ \\
				\ \ \ \ \begin{tabular}{l}
					$\Let{x}{\Load{b}{\scx}}{}$\\
					$\Store{b}{\scx}{v};$\\
					$\Store{a}{\scx}{x+(\ishigh\ ?\ 2:1)};$\\
					$\mathbf{repeat}$\\
					$\quad \Let{z}{\Load{a}{\scx}}{}$\\
					$\quad \quad \If{z==x+3}{1}{0}$\\
					$\mathbf{end}$\\
					$\Store{b}{\scx}{x+3};$\\
				\end{tabular}
			\end{minipage}
		\end{tabular}
		\ \\
		$\recv{d}\ \ \defeq$ \\
		\begin{tabular}{l}
			$\Let{t}{\Load{c}{\scx}}{}$\\
			$\mathbf{let}\ lv\ =\ \mathbf{(repeat}$\\
			$\qquad \qquad \qquad \Let{z}{\Load{a}{\scx}}{}$\\
			$\qquad \qquad \qquad \quad \If{(z==t+1)||(z==t+2)}{z}{0}$\\
			$\qquad \qquad \quad \ \mathbf{end)}\ \mathbf{in}$\\
			$\Let{v}{\Load{b}{\scx}}{}$\\
			$\Store{d}{\scx}{v};$\\
			$\Store{a}{\scx}{t+3};$\\
			$\Store{c}{\scx}{t+3};$\\
			$lv-t;$\\
		\end{tabular}
	\end{minipage}
	\caption{The sequentially-consistent synchronous channel implementation (baseline).\label{fig:baseline}}
\end{figure}
\fi 

\end{document}